\documentclass[aps,pre,twocolumn,10pt,superscriptaddress,nofootinbib,balancelastpage]{revtex4-1} 
\usepackage[latin1]{inputenc}
\usepackage{amsmath,amssymb}
\usepackage{mathrsfs} 
\usepackage[capitalise]{cleveref}
\usepackage{siunitx}
\usepackage{braket}
\usepackage{calc}
\usepackage{tabularx}
\usepackage{dsfont}
\usepackage{color}
\usepackage{ifthen}
\usepackage{graphicx}
\usepackage{seqsplit}

\allowdisplaybreaks

\newcommand{\Eins}{\mathds{1}}%

\newcommand{\ii}{\mathrm{i}}%

\newcommand{\dif}{\mathrm{d}}%
\newcommand{\tdif}[2]{\frac{\dif#1}{\dif#2}}%
\newcommand{\Nabla}{\vec{\nabla}}%

\newcommand{\norm}[1]{\lVert#1\rVert}%

\newcommand{\Tr}{\operatorname{Tr}}%
\newcommand{\ZT}[1]{\textquotedblleft#1\textquotedblright}%

\newcolumntype{Y}{>{\centering\arraybackslash}X}%
\newcolumntype{Z}{>{\raggedright\arraybackslash}X}%

\newlength{\myl}%
\newcommand{\SUM}[2]{{\setlength{\myl}{\widthof{$\displaystyle\sum_{#1}^{#2}$}*\real{0.5}-\widthof{$\displaystyle\sum$}*\real{0.5}}\sum_{#1}^{#2}\;\hspace{-\the\myl}}}
\newcommand{\INT}[3]{\settowidth{\myl}{$\displaystyle\int_{#1}^{#2}$}{\int_{#1}^{#2}\;\;\;\hspace{-\the\myl}\dif #3}\,}
\newcommand{\TINT}[3]{\settowidth{\myl}{$\int_{#1}^{#2}$}{\int_{#1}^{#2}\!\ifthenelse{\equal{#1#2}{}}{}{\;\;\;\;\hspace{-\the\myl}}\dif #3}\,}%
\newcommand{\EINT}[3]{\settowidth{\myl}{$\int_{#1}^{#2}$}{\int_{#1}^{#2}\;\;\;\,\hspace{-\the\myl}\dif #3}\,}

\begin{document}
\title{Master equations for Wigner functions with spontaneous collapse and their relation to thermodynamic irreversibility}

\author{Michael te Vrugt}
\affiliation{Institut f\"ur Theoretische Physik, Center for Soft Nanoscience, Westf\"alische Wilhelms-Universit\"at M\"unster, D-48149 M\"unster, Germany}
\affiliation{Philosophisches Seminar, Westf\"alische Wilhelms-Universit\"at M\"unster, D-48143 M\"unster, Germany}

\author{Gyula I. T\'{o}th}
\affiliation{Interdisciplinary Centre for Mathematical Modelling and Department of Mathematical Sciences, Loughborough University, Loughborough, LE11 3TU, United Kingdom}

\author{Raphael Wittkowski}
\email[Corresponding author: ]{raphael.wittkowski@uni-muenster.de}
\affiliation{Institut f\"ur Theoretische Physik, Center for Soft Nanoscience, Westf\"alische Wilhelms-Universit\"at M\"unster, D-48149 M\"unster, Germany}

\begin{abstract}
Wigner functions, allowing for a reformulation of quantum mechanics in phase space, are of central importance for the study of the quantum-classical transition. A full understanding of the quantum-classical transition, however, also requires an explanation for the absence of macroscopic superpositions to solve the quantum measurement problem. Stochastic reformulations of quantum mechanics based on spontaneous collapses of the wavefunction are a popular approach to this issue. In this article, we derive the dynamic equations for the four most important spontaneous collapse models -- Ghirardi-Rimini-Weber (GRW) theory, continuous spontaneous localization (CSL) model, Di{\'o}si-Penrose model, and dissipative GRW model -- in the Wigner framework. The resulting master equations are approximated by Fokker-Planck equations. Moreover, we use the phase-space form of GRW theory to test, via molecular dynamics simulations, David Albert's suggestion that the stochasticity induced by spontaneous collapses is responsible for the emergence of thermodynamic irreversibility. The simulations show that, for initial conditions leading to anti-thermodynamic behavior in the classical case, GRW-type perturbations do not lead to thermodynamic behavior. Consequently, the GRW-based equilibration mechanism proposed by Albert is not observed.
\end{abstract}
\maketitle

\section{Introduction}
Wigner functions provide a description of quantum systems in phase space. Since their initial development \cite{Wigner1932}, they have found a significant number of applications \cite{WeinbubF2018,FerryNWBWS2020,FerryN2018,RundleE2021} in fields such as atomic physics \cite{WeissCGB2016}, quantum optics \cite{SchulteHJMLA2015,Schleich2011,Olivares2012}, visualization of quantum effects \cite{RundleDDTE2020,FerraroFGTBFGD2013}, computational electronics \cite{KimKT2017,FerryW2018,SpisakWS2015}, and solid-state theory \cite{GrollHMWK2021,HahnWK2020,HahnGKW2019,WiggerGARK2016,JacoboniB2004}. Besides these practical aspects, they are also of interest for fundamental questions in quantum mechanics such as the theory of quantum chaos \cite{ZurekP1995}. Moreover, they have a natural connection to statistical mechanics \cite{CancellieriBJ2007,RundleTSDBE2019}, including nonequilibrium descriptions such as the Boltzmann equation \cite{NedjalkovWBSDF2019,Spohn2006,SelsB2013} and the Fokker-Planck equation \cite{Bondarev1992,JungelLM2011,ArnoldLMS2004}. In particular, it has been suggested that Wigner functions might be useful in understanding the origin of thermodynamic irreversibility \cite{Wallace2016}.

Since they allow to describe quantum mechanics in a way that has strong formal analogies to classical mechanics (Wigner functions are defined on phase space and governed by a dynamic equation that reduces to the classical Liouville equation for the phase-space density in the limit $\hbar \to 0$), an important application of Wigner functions is the study of the quantum-classical transition \cite{HabibJMRSS2002,JasiakMHH2009}. Wigner functions allow to develop a connection between classical and quantum models for liquid crystals \cite{teVrugtW2019c} and to study mixed quantum-classical dynamics \cite{teVrugtLW2020,BurghardtP2004,BurghardtB2006}. However, recovering the Liouville equation is not sufficient for recovering classical mechanics. Quantum systems can be in superpositions, whereas macroscopic superpositions (such as a cat being dead and alive at the same time \cite{Schroedinger1935}) are never observed. This is commonly attributed to a \ZT{collapse of the wavefunction} that takes place when a measurement is performed. 
However, this approach is often seen as unsatisfactory given that such a collapse is not allowed by the Schr\"odinger equation, meaning that one has to make the strange assumption that the Schr\"odinger equation does not apply to measurements. This issue is known as the \textit{measurement problem} \cite{FriebeKLNPS2018}. 

A popular approach to the measurement problem is to propose a stochastic modification of quantum mechanics in which collapses occur spontaneously \cite{GhirardiRW1986,GhirardiPR1990,GhirardiB2020,GhirardiGB1995}. Spontaneous collapse models make empirical predictions that differ slightly from those of \ZT{standard} quantum mechanics and can thus be tested experimentally \cite{BassiLSSU2013,Adler2007}. Such experiments have become an important field of research in recent years \cite{VinanteCBCVFMMU2020,DonadiPCDLB2020,ZhengEtAl2020}. Incorporating such spontaneous collapses in the theory of Wigner functions is desirable for at least four reasons: First, it allows to solve the problem of the classical-quantum transition in the Wigner framework also regarding the measurement problem. Second, if collapse models should be confirmed experimentally, the Wigner formalism should of course allow to describe them in order to remain a powerful alternative to Hilbert-space quantum mechanics. Third, Wigner functions are very useful for the visualization of collapse effects, as has been exploited for both spontaneous \cite{SchrinskiSH2017,DasLSS2013} and environment-induced \cite{GarrawayK1994} collapses. Fourth, modern quantum technology -- a field where Wigner functions are of high importance \cite{WeinbubF2018} -- may be relevant for experimental tests of spontaneous collapse models. For example, it has been discussed whether SQUIDs (superconducting quantum interference  devices) can be used for such a test \cite{BuffaNR1995,Rae1990}.

In addition, spontaneous collapse models might provide a solution to another fundamental problem of physics, namely to the aforementioned problem of thermodynamic irreversibility \cite{Toth2020,teVrugt2020}. This problem (which actually includes a variety of subproblems \cite{teVrugt2020}) is concerned with the fact that macroscopic thermodynamics has a clear arrow of time associated with the monotonous increase of entropy, whereas the microscopic laws of (classical or quantum) mechanics are invariant under time reversal. David Albert \cite{Albert1994,Albert1994b,Albert2000} has suggested that spontaneous collapses of the wavefunction might be responsible for the irreversible approach to equilibrium. Although it has received significant attention in philosophy of physics \cite{HemmoS2001,HemmoS2003,HemmoS2005,Price2002,Monton2004,North2011}, rigorous physical tests of this proposal are lacking at present.

In this work, we provide a reformulation of the four most important spontaneous collapse models, the Ghirardi-Rimini-Weber (GRW) theory \cite{GhirardiRW1986}, the continuous spontaneous localization (CSL) model \cite{Pearle1989,GhirardiPR1990}, the Di{\'o}si-Penrose model \cite{Penrose2014,Penrose1996,Diosi1987,Diosi1989}, and the dissipative GRW model \cite{SmirneVB2014,SmirneB2015}, based on Wigner functions. We then provide an analytical and numerical discussion of Albert's proposal, thereby investigating whether GRW theory might actually contribute to the solution of the problem of irreversibility. The mechanism suggested by Albert (deviations resulting from stochastic perturbations lead from anti-thermodynamic to thermodynamic behavior) is not observed in the simulations.

This article consists of two main parts. Although both contain important results, a reader only interested in one of them may skip the other one. First, we address the reformulation of collapse models in the Wigner framework: In \cref{wignerfunctions}, we provide a brief introduction to Wigner functions. We then apply the Wigner transformation to the GRW theory (\cref{grwtheory}), CSL model (\cref{cslmodel}), Di{\'o}si-Penrose model (\cref{diosipenrosemodel}), and dissipative GRW theory (\cref{dgm}). Second, we discuss Albert's GRW-based explanation of irreversibility: The connection of GRW collapses to thermodynamic irreversibility is analyzed in \cref{irre}. In \cref{deco}, decoherence-based alternatives are discussed. A numerical test of the GRW-based explanation of irreversibility is provided in \cref{numeric}. We conclude in \cref{conclusion}.

\section{\label{wignerfunctions}Wigner functions}
In \ZT{standard} quantum mechanics, a system is described by the density operator\footnote{Throughout this work, we use a hat $\hat{}$ to denote a quantum-mechanical operator.} (also known as \ZT{density matrix} or \ZT{statistical operator}) $\hat{\rho}$ whose time evolution is given by the Liouville-von-Neumann equation 
\begin{equation}
\tdif{}{t}\hat{\rho}= -\frac{\ii}{\hbar}[\hat{H},\hat{\rho}]
\label{lvn}
\end{equation}
with time $t$, imaginary unit $\ii$, reduced Planck constant $\hbar$, commutator $[\cdot,\cdot]$, and Hamiltonian $\hat{H}$. Furthermore, an observable is represented by a self-adjoint operator $\hat{A}$.

An alternative to the density-operator-based formalism is to use \textit{Wigner functions} \cite{Wigner1932}. Here, the state of the system is described by a function $W(x,p)$ (Wigner function) that depends on the phase-space coordinates $x$ and $p$. (In this work, we only require Wigner functions depending on position $x$ and momentum $p$. However, the approach is much more general and can also be applied to systems with other degrees of freedom, such as spin \cite{TilmaESMN2016,teVrugtW2019c}.) The Wigner function is a generalization of a classical phase-space distribution function, from which it differs in that it can also take negative values \cite{Case2008}. For a one-particle system in one dimension, the Wigner function is defined as \cite{GneitingFH2013}
\begin{equation}
W(x,p) = \frac{1}{2\pi\hbar}\INT{}{}{x'}\bigg\langle{x-\frac{1}{2}x'
\bigg\lvert\hat{\rho}\bigg\rvert
x+\frac{1}{2}x'\bigg\rangle e^{\ii\frac{x' p}{\hbar}}}.
\label{Wigner}%
\end{equation}
Throughout this work, the dependence of the Wigner function $W$ and density operator $\hat{\rho}$ on time $t$ is not written explicitly to improve readability. For $N$ particles in three dimensions, \cref{Wigner} generalizes to \cite{HilleryOSW1984}
\begin{align}
&W(\vec{r}_1,\dotsc,\vec{r}_N,\vec{p}_1,\dotsc,\vec{p}_N) \notag\\ 
&= \frac{1}{(2\pi\hbar)^{3N}}\INT{}{}{^3r_1'}\dotsi\INT{}{}{^3r_N'} e^{\frac{\ii}{\hbar}\sum_{i=1}^{N}\vec{r}_i' \cdot\vec{p}_i} \label{Wignermany}\\
&\quad\, \bigg\langle{\vec{r}_1-\frac{1}{2}\vec{r}_1',\dotsc,\vec{r}_N - \frac{1}{2}\vec{r}_N'
\bigg\lvert\hat{\rho}\bigg\rvert
\vec{r}_1+\frac{1}{2}\vec{r}_1',\dotsc, \vec{r}_N+\frac{1}{2}\vec{r}_N'\bigg\rangle}. \notag
\end{align}
Equations \eqref{Wigner} and \eqref{Wignermany} can not only be applied to the density operator $\hat{\rho}$, but to an arbitrary operator $\hat{A}$. This gives its so-called \textit{Weyl symbol} $A(x,p)$.\footnote{In the convention used here, which is identical to the one in Ref.\ \cite{GneitingFH2013}, the equation for $A(x,p)$ is given by \cref{Wigner} with $W$ replaced by $A$, $\hat{\rho}$ replaced by $\hat{A}$, and the prefactor $(2\pi\hbar)^{-1}$ removed. See Ref.\ \cite{teVrugtW2019c} for a discussion of the prefactor and a different convention.} The expectation value $\braket{\hat{A}}$ of the observable $\hat{A}$, which in \ZT{standard} quantum mechanics is given by
\begin{equation}
\braket{\hat{A}} = \Tr(\hat{A}\hat{\rho})
\end{equation}
with the quantum-mechanical trace $\Tr$, can then be calculated as \cite{GneitingFH2013}
\begin{equation}
\braket{\hat{A}} = \INT{}{}{x}\INT{}{}{p}A(x,p)W(x,p)    
\end{equation}
just as in classical mechanics.

We now discuss the dynamics of Wigner functions. First, we introduce the \textit{star product} \cite{Groenewold1946}: When transforming to phase space, one has to replace the product $\hat{A}\hat{B}$ of two Hilbert space operators $\hat{A}$ and $\hat{B}$ by their star product $A \star B$, which is defined as \cite{Fairlie1999}
\begin{equation}
A(x,p)\star B(x,p) = A(x,p)\exp\!\bigg(\frac{\ii \hbar}{2}(\overleftarrow{\partial_x}\overrightarrow{\partial_p}-\overleftarrow{\partial_p}\overrightarrow{\partial_x})\bigg)B(x,p).
\label{starproduct}
\end{equation}
Here, the arrows indicate in which direction the derivatives act. The star product \eqref{starproduct} allows to introduce the \textit{Moyal bracket} \cite{Moyal1949,Groenewold1946} 
\begin{equation}
\begin{split}
&\{A(x,p),B(x,p)\}_\star \\
&= \frac{\ii}{\hbar}(A(x,p) \star B(x,p) - B(x,p)\star A(x,p))\\
&= \frac{2}{\hbar} A(x,p) \sin\!\bigg(\frac{ \hbar}{2}(\overleftarrow{\partial_x}\overrightarrow{\partial_p}-\overleftarrow{\partial_p}\overrightarrow{\partial_x})\bigg)B(x,p),
\end{split}    
\label{moyalbracket}
\end{equation}
which is the Wigner equivalent of the commutator.
When neglecting terms of order $\hbar^2$, the Moyal bracket \eqref{moyalbracket} reduces to the classical Poisson bracket. Combining \cref{lvn,moyalbracket} and taking into account the correspondence between commutators and Moyal brackets leads to the dynamic equation for the time evolution of Wigner functions
\begin{equation}
\partial_t W(x,p) = \{H(x,p),W(x,p)\}_\star,  
\label{lvnwigner}
\end{equation}
where $H$ is the Weyl symbol of the Hamiltonian $\hat{H}$. This Weyl symbol is given by $H(x,p) = p^2/(2m) + U(x)$ with the particle mass $m$ and potential $U$ \cite{Case2008}. Equation \eqref{lvnwigner} reduces to the Liouville equation of classical mechanics if terms of order $\hbar$ are neglected or if the potential is at most a second-order polynomial in $x$ \cite{GneitingFH2013}. (Some difficulties of the limit $\hbar \to 0$ are discussed in Ref.\ \cite{Case2008}. In particular, the classical limit cannot be taken in this simple way if the Hamiltonian generates chaotic classical dynamics \cite{JacquodP2009}.) An overview over theory and applications of Wigner functions is given in Refs.\ \cite{FerryN2018,WeinbubF2018,HilleryOSW1984,Lee1995,RundleE2021}.

\section{\label{grwtheory}GRW theory}
The formalism introduced in \cref{wignerfunctions} is of interest for the \textit{Ghirardi-Rimini-Weber} (GRW) theory  \cite{GhirardiRW1986}, which is a modified form of quantum mechanics. It allows to deal with the famous quantum-mechanical \textit{measurement problem} \cite{Schlosshauer2005,Wigner1963,FriebeKLNPS2018,Wallace2012,Albert1992}: If one performs a measurement on a quantum system that is in a superposition, then, assuming that both system and measurement apparatus obey the (linear, deterministic, and unitary) Schr\"odinger equation, the apparatus will, at the end, be in a superposition too. The quantum-mechanical time evolution inevitably leads to macroscopic superpositions, as is well known from the thought experiment on Schr\"odinger's cat \cite{Schroedinger1935}. However, this is not what we observe. A measurement on a system in a superposition produces a definite outcome. Therefore, one typically postulates a \ZT{collapse of the wavefunction}, which is a sudden transition to an eigenstate of the observable one has measured as a consequence of measurement. This postulate, however, has no basis in the deterministic Schr\"odinger equation. A variety of solutions to this problem have been proposed, including hidden variables (\ZT{Bohmian mechanics} \cite{Bohm1952,Bohm1952b}), superdeterminism \cite{HossenfelderP2020,Hossenfelder2020}, modal interpretations \cite{LombardiD2017}, and the existence of many worlds in which all outcomes are realized (\ZT{Everett interpretation}, also known as \ZT{many-worlds interpretation} \cite{Everett1957,Wallace2012}). It is sometimes claimed that decoherence effects (see \cref{deco}) provide a solution to the measurement problem. This, however, is wrong, since decoherence alone does not determine which one of two possible measurement outcomes occurs \cite{Schlosshauer2005,Adler2003}.

The GRW theory \cite{GhirardiRW1986} (reviewed in Refs.\ \cite{BassiG2003,GhirardiB2020,BassiLSSU2013}) takes a different approach by assuming that the collapse of the wavefunction is not a consequence of measurements, but something that happens spontaneously. For a system with $N$ particles, the deterministic dynamics given by \cref{lvn} is replaced by the more general master equation \cite{GhirardiRW1986}
\begin{equation}
\tdif{}{t}\hat{\rho}= - \frac{\ii}{\hbar}[\hat{H},\hat{\rho}]-\sum_{i=1}^{N}\lambda_i(\hat{\rho}-\hat{T}_i(\hat{\rho})),
\label{grw}
\end{equation}
where $\lambda_i$ is the localization rate for the $i$-th particle. In this work, we assume that $\lambda_i = \lambda$ with a constant $\lambda$ for all particles. Since the localization rate is often assumed to depend on the particle mass \cite{PearleS1994,FeldmannT2012}, this can be motivated by the assumption that the particles belong to the same species. $\hat{T}_i$ is the operator
\begin{equation}
\hat{T}_i(\hat{\rho})= \sqrt{\frac{\alpha}{\pi}}\INT{-\infty}{\infty}{x}
e^{-\frac{\alpha}{2}(\hat{x}_i - x)^2}\hat{\rho}\,  
e^{-\frac{\alpha}{2}(\hat{x}_i - x)^2}  
\label{standardcollapseoperator}
\end{equation}
with the position operator for the $i$-th particle $\hat{x}_i$ and a constant $\alpha$ determining the width of the collapsed function. It is easily shown that \cref{grw} is trace-preserving \cite{GhirardiRW1986}. The idea of Eq.\ \eqref{grw} is that \ZT{spontaneous collapse of the wavefunction} simply means that the particle's wavefunction is, at a certain rate $\lambda$, stochastically multiplied by a Gaussian. The parameters are tuned in such a way that these collapses are very rare for small systems (such as quantum particles), but very common for large systems (such as a measurement apparatus). This explains why superpositions are observed in microscopic, but not in macroscopic systems.

The GRW theory (as well as other approaches to the measurement problem) is conceptually very important for a particular application of Wigner functions, namely the study of the classical limit. As discussed in \cref{wignerfunctions}, the governing equation \eqref{lvnwigner} for Wigner functions reduces to the classical Liouville equation for $\hbar \to 0$. The measurement problem shows that recovering the Liouville equation is not sufficient to recover the observations of classical mechanics: If a cat is in a superposition of \ZT{being alive} and \ZT{being dead}, the Liouville dynamics will not change this (a superposition of a dead and a living cat evolving classically remains a superposition of a dead and a living cat). What we require is a mechanism that explains why such superpositions are not observed in the macroscopic world. The GRW theory is one such mechanism. 

A typical problem of spontaneous collapse models is that energy is not conserved. In fact, due to the noise, it increases at a constant rate. Although this effect is extremely small (the temperature of an ideal GRW gas increases by about $10^{-15}$ K per year \cite{GhirardiRW1986}), it is still an undesirable feature. To solve this problem, dissipative extensions of the GRW theory \cite{SmirneVB2014}, CSL model \cite{SmirneB2015}, and Di{\'o}si-Penrose model \cite{BahramiSB2014} have been derived. They ensure that the energy remains finite during the time evolution of the system. These approaches are discussed in \cref{dgm}. Further extensions of GRW theory (not considered in this work) are models with colored rather than white noise for the stochastic perturbations of the wavefunction \cite{BassiG2002} and relativistic extensions \cite{Tumulka2006}.

We now transform the governing equation \eqref{grw} of the GRW theory to the Wigner formalism. We first demonstrate this derivation for the case of one particle in one spatial dimension to keep the notation compact, and then generalize to the case of $N$ particles in three dimensions which is more relevant for statistical mechanics.

First, we use the fact that \cite{GhirardiRW1986}
\begin{equation}
\braket{x'|\hat{T}(\hat{\rho})|x''} = e^{-\frac{\alpha}{4}(x'-x'')^2}\braket{x'|\hat{\rho}|x''},
\label{fact}
\end{equation}
where $\ket{x}$ is a position eigenstate with eigenvalue $x$. The operator $\hat{T}(\hat{\rho})$ is subject to the same transformation rule \eqref{Wigner} as the density operator $\hat{\rho}$ itself. Using \cref{fact}, we find
\begin{widetext}
\begin{align}
T(W(x,p))&=\frac{1}{2\pi\hbar}\INT{}{}{x'}\bigg\langle{x-\frac{1}{2}x'\bigg\lvert \hat{T}(\hat{\rho})\bigg\rvert x+\frac{1}{2}x'\bigg\rangle e^{\ii\frac{x' p}{\hbar}}}\notag\\
&= \sqrt{\frac{\alpha}{\pi}}\INT{}{}{x}\frac{1}{2\pi\hbar}\INT{}{}{x'}\bigg\langle{x-\frac{1}{2}x'\bigg\lvert e^{-\frac{\alpha}{2}(\hat{x} - x)^2}\hat{\rho}\, e^{-\frac{\alpha}{2}(\hat{x} - x)^2}\bigg\rvert x+\frac{1}{2}x'\bigg\rangle e^{\ii\frac{x' p}{\hbar}}}\notag\\
&=
\frac{1}{2\pi\hbar}\INT{}{}{x'}e^{-\frac{\alpha}{4}((x-\frac{1}{2}x')-(x+\frac{1}{2}x'))^2}
\bigg\langle{x-\frac{1}{2}x' \bigg\lvert\hat{\rho} \bigg\rvert x+\frac{1}{2}x'\bigg\rangle e^{\ii\frac{x' p}{\hbar}}}\notag\\
&=
\frac{1}{2\pi\hbar}\INT{}{}{x'}e^{-\frac{\alpha}{4}(x')^2}
\bigg\langle{x-\frac{1}{2}x'\bigg\lvert\hat{\rho} \bigg\rvert x+\frac{1}{2}x'\bigg\rangle e^{\ii\frac{x' p}{\hbar}}}\label{ttransformation}\\
&=
\frac{1}{2\pi\hbar}\sqrt{\frac{1}{\pi\alpha\hbar^2}}\INT{}{}{x'}\INT{}{}{p'}e^{-\frac{1}{\alpha\hbar^2}(p')^2} e^{-\ii\frac{x' p'}{\hbar}}
\bigg\langle{x-\frac{1}{2}x'\bigg\lvert \hat{\rho} \bigg\rvert x+\frac{1}{2}x'\bigg\rangle e^{\ii\frac{x' p}{\hbar}}}\notag\\
&=
\frac{1}{2\pi\hbar}\sqrt{\frac{1}{\pi\alpha\hbar^2}}\INT{}{}{x'}\INT{}{}{p'}e^{-\frac{1}{\alpha\hbar^2}(p')^2}
\bigg\langle{x-\frac{1}{2}x'\bigg\lvert \hat{\rho} \bigg\rvert x+\frac{1}{2}x'\bigg\rangle e^{\ii\frac{x' (p-p')}{\hbar}}}\notag\\
&=
\frac{1}{\sqrt{\pi\alpha\hbar^2}}\INT{}{}{p'}e^{-\frac{1}{\alpha\hbar^2}(p')^2}W(x,p-p').\notag
\end{align}

Consequently, the spontaneous multiplication by a Gaussian function of the position mediated by the operator $\hat{T}(\hat{\rho})$ in the Hilbert space formalism corresponds to a convolution with a Gaussian function of the momentum in the Wigner function formalism.

We can now combine \cref{grw,lvnwigner,ttransformation} to the \textit{GRW master equation for Wigner functions}, given by
\begin{equation}
\partial_t W(x,p) = \{H(x,p),W(x,p)\}_\star  - \lambda \bigg(W(x,p) - \frac{1}{\sqrt{\pi\alpha\hbar^2}}\INT{}{}{p'}e^{-\frac{1}{\alpha\hbar^2}(p')^2}W(x,p-p')\bigg).
\label{grwwigner}
\end{equation}
To simplify \cref{grwwigner}, we make two approximations. First, we use a Kramers-Moyal expansion \cite{Hanggi1984}, which at second order leads to a Fokker-Planck equation \cite{GaveauMT1997}. For this purpose, we Taylor expand $W(x,p-p')$ around $p'=0$ and find
\begin{equation}
\partial_t W(x,p) = \{H(x,p),W(x,p)\}_\star +\sum_{n=1}^{\infty} \lambda\frac{(\alpha\hbar^2)^n(2n-1)!!}{2^n (2n)!}\partial_p^{2n}W(x,p)
\label{kramersmoyal}
\end{equation}
where ! is the factorial and !! is the double factorial. When truncating at order $n=1$, we obtain from \cref{kramersmoyal} the Fokker-Planck equation
\begin{equation}
\partial_t W(x,p) = \{H(x,p),W(x,p)\}_\star + D_p \partial_p^2 W(x,p)
\label{momentumdiffusion}
\end{equation}
with the momentum diffusion coefficient $D_p = \lambda\alpha\hbar^2/4$. Equations of the same form (but with a different physical content) have been obtained in the context of decoherence \cite{Zurek2003}. Here, one typically also has a friction term (see \cref{deco}), \cref{momentumdiffusion} is then obtained when taking the limit of vanishing friction at fixed diffusion constant \cite{ZurekP1995}. Moreover, equations similar to \cref{momentumdiffusion} have been obtained for the collapse-induced \cite{SchrinskiSH2017} and decoherence-induced \cite{SticklerSH2018} dynamics of quantum rotors.

Since \cref{momentumdiffusion} is of order $\hbar^2$, the first term in \cref{grwwigner} (the Moyal bracket) should also be expanded up to this order. This gives
\begin{equation}
\{H(x,p),W(x,p)\}_\star \approx - \frac{p}{m}\partial_x W(x,p) + (\partial_x U(x)) (\partial_p W(x,p)) - \frac{\hbar^2}{24}(\partial_x^3 U(x)) (\partial_p^3 W(x,p)).
\label{moyalbracketexpansion}
\end{equation}
We have assumed here that $H$ has the form $H(x,p) = p^2/(2m) + U(x)$ with a potential $U$. If $U$ is quadratic, linear, or constant, e.g., if we have a free particle or a harmonic oscillator, the time evolution obtained from \cref{moyalbracketexpansion} is thus classical. Making this assumption, the Fokker-Planck equation \eqref{momentumdiffusion} reduces to
\begin{equation}
\partial_t W(x,p) = - \frac{p}{m}\partial_x W(x,p) + (\partial_x U(x)) (\partial_p W(x,p))  + D_p \partial_p^2 W(x,p).
\label{grwfokker}
\end{equation}
The same result is obtained if, e.g., due to a weak potential, the last term in \cref{moyalbracketexpansion} is small compared to the diffusion term in \cref{grwfokker}.

Now, we consider $N$ particles in three dimensions. A calculation analogous to \cref{ttransformation} gives
\begin{equation}
\sum_{i=1}^{N}T_i(W(\{\vec{r}_j,\vec{p}_j\})) = \sum_{i=1}^{N} \frac{1}{(\pi\alpha\hbar^2)^{\frac{3}{2}}}\INT{}{}{^3 p'}e^{-\frac{1}{\alpha\hbar^2}(\vec{p}')^2} W(\{\vec{r}_j\},\vec{p}_1,\dotsc,\vec{p}_{i-1},\vec{p}_i-\vec{p}',\vec{p}_{i+1},\dotsc).  \label{grwwignermany}
\end{equation}
Hence, the extension of \cref{grwwigner} to $N$ particles in three dimensions is given by
\begin{equation}
\partial_t W(\{\vec{r}_j,\vec{p}_j\}) = \{H(\{\vec{r}_j,\vec{p}_j\}), W(\{\vec{r}_j,\vec{p}_j\})\}_\star -\lambda\bigg(W(\{\vec{r}_j,\vec{p}_j\}) - \sum_{i=1}^{N}T_i(W(\{\vec{r}_j,\vec{p}_j\}))\bigg) 
\label{generalgrwmodel}
\end{equation}
with $\sum_i T_i$ given by \cref{grwwignermany}.
The Fokker-Planck equation \eqref{grwfokker} generalizes to
\begin{equation}
\partial_t W(\{\vec{r}_j,\vec{p}_j\}) = \sum_{i=1}^{N}\bigg(- \frac{\vec{p}_i}{m}\cdot\Nabla_{\vec{r}_i}W(\{\vec{r}_j,\vec{p}_j\}) + (\Nabla_{\vec{r}_i} U(\{\vec{r}_j\}))\cdot(\Nabla_{\vec{p}_i} W(\{\vec{r}_j,\vec{p}_j\})) + D_p \Nabla_{\vec{p}_i}^2 W(\{\vec{r}_j,\vec{p}_j\})\bigg)
\label{grwfokkermany}
\end{equation}
\end{widetext}
with $D_p = 3\lambda\alpha\hbar^2/4$. 
This derivation shows that the Wigner approach is very useful in the study of spontaneous collapse models: Using Hilbert space theory, the derivation of a Fokker-Planck equation for GRW theory (as done in Ref.\ \cite{GhirardiRW1986}) requires several pages of calculation and a significant number of auxiliary assumptions. In the Wigner formalism, however, it is obtained almost immediately and in a very natural way. On the other hand, it is notable that \cref{grwwigner} has the form of a convolution in momentum space, whereas in \cref{grw} the density operator is simply multiplied by a Gaussian in position space. For this reason, if we do not operate with a simplified (Fokker-Planck) model, it is easier to work with the Fourier-transformed Wigner function
\begin{equation}
\tilde{W}(x,x') = \INT{}{}{p}W(x,p)e^{-\ii\frac{x'p}{\hbar}} 
\label{transformedwigner}
\end{equation}
that depends on two position coordinates rather than on a position and a momentum coordinate. A physical reason for why it is easier to work with the Fourier-transformed Wigner function \eqref{transformedwigner} is that Wigner functions typically treat position and momentum on an equal footage, whereas in GRW theory collapses only take place in position space. For the Wigner function \eqref{transformedwigner}, \cref{ttransformation} simplifies to
\begin{equation}
T(\tilde{W}(x,x'))= e^{-\frac{\alpha}{4}(x')^2}\tilde{W}(x,x'). 
\label{fact2}
\end{equation}
The result \eqref{fact2} also directly follows from \cref{fact} together with the fact that $\tilde{W}$ is related to the density operator in position representation, $\rho(x,x')=\braket{x|\hat{\rho}|x'}$, by $\tilde{W}(x,x')=\rho(x - x'/2, x+ x'/2)$ (as is obvious from \cref{Wigner}). The relation of $W$ and $\rho$ via the Fourier transformation \cite{Frensley1987} also explains why we need a convolution when describing the GRW dynamics in terms of Wigner functions (a Fourier transformation converts a multiplication to a convolution).

\section{\label{cslmodel}CSL model}
A drawback of the GRW theory is that it does not preserve the symmetry properties of a system of identical particles \cite{BassiG2003}. In particular, it does not respect the fact that the wavefunction of a system of identical particles has to be symmetric under particle exchange for bosons and antisymmetric for fermions. This problem was solved by the continuous spontaneous localization (CSL) model introduced in Refs.\ \cite{Pearle1989,GhirardiPR1990}. We here follow Ref.\ \cite{BassiG2003}. First, we define the locally averaged density operator
\begin{equation}
\hat{N}(\vec{r})=\sum_{s}^{}\INT{}{}{^3r'}g(\vec{r}'-\vec{r})\hat{a}^\dagger (\vec{r}',s)\hat{a} (\vec{r}',s),   
\end{equation}
where $\hat{a}^\dagger (\vec{r}',s)$ and $\hat{a}^\dagger (\vec{r}',s)$ are creation and annihilation operators for a particle with spin component $s$ at position $\vec{r}'$ and $g$ is a symmetric positive function typically chosen to be
\begin{equation}
g(\vec{r})=\Big(\frac{\alpha}{2\pi}\Big)^\frac{3}{2}\exp\!\Big(-\frac{\alpha}{2} \vec{r}^2\Big).    
\end{equation}
The density operator is assumed to follow the dynamic equation
\begin{align}
\tdif{}{t}\hat{\rho} &= - \frac{\ii}{\hbar}[\hat{H},\hat{\rho}] \notag\\ 
&\quad\,+\xi\INT{}{}{^3r}\Big(\hat{N}(\vec{r})\hat{\rho}\hat{N}(\vec{r}) -\frac{1}{2}[ \hat{N}^2(\vec{r}),\hat{\rho}]_+\Big)
\label{csl}
\end{align}
with a constant $\xi$ that is related to the localization rate, see \cref{lambdaxi} below, and the anticommutator $[\cdot,\cdot]_+$.  From here on, we ignore the spin degrees of freedom since they are not relevant for the collapse dynamics. See Ref.\ \cite{BassiG2003} for a discussion of how \cref{csl} follows from the theory of stochastic processes in Hilbert space.

\begin{widetext}
We now derive the Wigner representation of \cref{csl}. In position representation, the last two terms on the right-hand side of \cref{csl} read \cite{BassiG2003}
\begin{equation}
\frac{\xi}{2}\bigg(\frac{\alpha}{4\pi}\bigg)^{\frac{3}{2}}\sum_{i,j=1}^{N}\braket{\vec{r}_1,\dotsc,\vec{r}_N|\hat{\rho}|\vec{r}_1',\dotsc,\vec{r}_N'}
\Big(2e^{-\frac{\alpha}{4}(\vec{r}_i-\vec{r}_j')^2}-e^{-\frac{\alpha}{4}(\vec{r}_i-\vec{r}_j)^2}-e^{-\frac{\alpha}{4}(\vec{r}_i'-\vec{r}_j')^2}\Big).
\end{equation}
Thus, the Weyl symbol for the operator given by the last two terms on the right-hand side of \cref{csl} is given by
\begin{align}
&\frac{1}{(2\pi\hbar)^{3N}}\INT{}{}{^3r_1'}\dotsb\INT{}{}{^3r_N'}\frac{\xi}{2}\bigg(\frac{\alpha}{4\pi}\bigg)^{\frac{3}{2}}\sum_{i,j=1}^{N}\bigg\langle\vec{r}_1-\frac{\vec{r}_1'}{2},\dotsc,\vec{r}_N-\frac{\vec{r}_N'}{2}
\bigg\lvert\hat{\rho}\bigg\rvert\vec{r}_1
+\frac{\vec{r}_1}{2},\dotsc,\vec{r}_N+\frac{\vec{r}_N'}{2}\bigg\rangle\notag\\
&\,e^{\frac{\ii}{\hbar}\sum_{k=1}^{N}\vec{r}_k'\cdot\vec{p}_k}
\Big(2e^{-\frac{\alpha}{4}(\vec{r}_i-\frac{1}{2}\vec{r}_i'-\vec{r}_j-\frac{1}{2}\vec{r}_j')^2}
-e^{-\frac{\alpha}{4}(\vec{r}_i-\frac{1}{2}\vec{r}_i'-\vec{r}_j+\frac{1}{2}\vec{r}_j')^2}
-e^{-\frac{\alpha}{4}(\vec{r}_i+\frac{1}{2}\vec{r}_i'-\vec{r}_j'-\frac{1}{2}\vec{r}_j')^2}\Big)\notag\\
&=\frac{1}{(2\pi\hbar)^{3N}}\bigg(\frac{1}{\pi\alpha\hbar^2}\bigg)^\frac{3}{2}\INT{}{}{^3r_1'}\dotsb\INT{}{}{^3r_N'}\INT{}{}{^3 p'}\frac{\xi}{2}\bigg(\frac{\alpha}{4\pi}\bigg)^{\frac{3}{2}}\sum_{i,j=1}^{N}\bigg\langle\vec{r}_1-\frac{\vec{r}_1'}{2},\dotsc,\vec{r}_N-\frac{\vec{r}_N'}{2}
\bigg\lvert\hat{\rho}\bigg\rvert\vec{r}_1
+\frac{\vec{r}_1}{2},\dotsc,\vec{r}_N+\frac{\vec{r}_N'}{2}\bigg\rangle\notag\\
&\quad\; e^{\frac{\ii}{\hbar}\sum_{k=1}^{N}\vec{r}_k'\cdot\vec{p}_k}e^{-\frac{(\vec{p}')^2}{\alpha\hbar^2}} \Big(2e^{\frac{\ii}{\hbar}(\vec{r}_i-\frac{1}{2}\vec{r}_i'-\vec{r}_j-\frac{1}{2}\vec{r}_j')\cdot\vec{p}'}
-e^{\frac{\ii}{\hbar}(\vec{r}_i-\frac{1}{2}\vec{r}_i'-\vec{r}_j+\frac{1}{2}\vec{r}_j')\cdot\vec{p}'}
-e^{\frac{\ii}{\hbar}(\vec{r}_i+\frac{1}{2}\vec{r}_i'-\vec{r}_j'-\frac{1}{2}\vec{r}_j')\cdot\vec{p}'}\Big)\\
&=\bigg(\frac{1}{\pi\alpha\hbar^2}\bigg)^\frac{3}{2}\INT{}{}{^3 p'}\frac{\xi}{2}\bigg(\frac{\alpha}{4\pi}\bigg)^{\frac{3}{2}}e^{-\frac{(\vec{p}')^2}{\alpha\hbar^2}}\sum_{i,j=1}^{N}e^{\frac{\ii}{\hbar}(\vec{r}_i-\vec{r}_j)\cdot\vec{p}'}\bigg(2W\bigg(\{\vec{r}_k\},\vec{p}_1,\dotsc,\vec{p}_i-\frac{\vec{p}'}{2},\vec{p}_j-\frac{\vec{p}'}{2},\dotsc,\vec{p}_N\bigg) \notag\\
&\quad\,-W\bigg(\{\vec{r}_k\},\vec{p}_1,\dotsc,\vec{p}_i-\frac{\vec{p}'}{2},\vec{p}_j+\frac{\vec{p}'}{2},\dotsc,\vec{p}_N\bigg) -W\bigg(\{\vec{r}_k\},\vec{p}_1,\dotsc,\vec{p}_i+\frac{\vec{p}'}{2},\vec{p}_j-\frac{\vec{p}'}{2},\dotsc,\vec{p}_N\bigg)\!\bigg).\notag
\end{align}
Therefore, the \textit{CSL master equation for Wigner functions} reads
\begin{align}
\partial_t W(\{\vec{r}_i,\vec{p}_i)\} &= \{H(\{\vec{r}_i,\vec{p}_i\}),W(\{\vec{r}_i,\vec{p}_i\}\}_\star \notag\\
&\quad\,+\bigg(\frac{1}{\pi\alpha\hbar^2}\bigg)^\frac{3}{2}\INT{}{}{^3 p'}\frac{\xi}{2}\bigg(\frac{\alpha}{4\pi}\bigg)^{\frac{3}{2}}e^{-\frac{(\vec{p}')^2}{\alpha\hbar^2}}\sum_{i,j=1}^{N}e^{\frac{\ii}{\hbar}(\vec{r}_i-\vec{r}_j)\cdot\vec{p}'}\bigg(2W\bigg(\{\vec{r}_k\},\vec{p}_1,\dotsc,\vec{p}_i-\frac{\vec{p}'}{2},\vec{p}_j-\frac{\vec{p}'}{2},\dotsc,\vec{p}_N\bigg) \notag\\
&\quad\,-W\bigg(\{\vec{r}_k\},\vec{p}_1,\dotsc,\vec{p}_i-\frac{\vec{p}'}{2},\vec{p}_j+\frac{\vec{p}'}{2},\dotsc,\vec{p}_N\bigg) -W\bigg(\{\vec{r}_k\},\vec{p}_1,\dotsc,\vec{p}_i+\frac{\vec{p}'}{2},\vec{p}_j-\frac{\vec{p}'}{2},\dotsc,\vec{p}_N\bigg)\!\bigg).
\label{cslwigner}
\end{align}
\end{widetext}
Equations \eqref{cslwigner} and \eqref{generalgrwmodel} coincide for $N=1$ if we set \cite{BassiG2003}
\begin{equation}
\lambda = \xi \bigg(\frac{\alpha}{4\pi}\bigg)^{\frac{3}{2}}.
\label{lambdaxi}
\end{equation}

\section{\label{diosipenrosemodel}Quantum gravity}
We now come back to the measurement problem. An interesting solution in the spirit of dynamical collapse models was proposed by Penrose  \cite{Penrose2014,Penrose1996} and Di{\'o}si \cite{Diosi1987,Diosi1989}, which is based on the problem of quantum gravity: As is well known, there is at present no generally accepted and experimentally verified unification of quantum mechanics and general relativity. While most approaches to this issue are based on quantizing gravity, an alternative approach consists in \ZT{gravitizing quantum mechanics} \cite{Penrose2014}, i.e., in modifying quantum mechanics to incorporate effects of gravity. A spatial quantum superposition generates a superposition of spacetimes, which is supposed to lead to a gravity-induced collapse of the wavefunction \cite{DonadiPCDLB2020}. Since this effect will depend on the mass of the system, it will be important for macroscopic, but negligible for microscopic systems; explaining why superpositions are not observed on macroscopic scales. Recent experimental tests \cite{DonadiPCDLB2020}, however, found no evidence of a gravity-induced collapse of the wave function, putting significant constraints on the possible validity of the Di{\'o}si-Penrose model.

A simple quantitative model for this effect is given by \cite{DonadiPCDLB2020,Diosi1987,Diosi1989} 
\begin{equation}
\tdif{}{t}\hat{\rho} = - \frac{\ii}{\hbar}[\hat{H},\hat{\rho}] - \frac{4\pi G}{\hbar}\INT{}{}{^3r}\INT{}{}{^3r'}\frac{[\hat{M}(\vec{r}'),[\hat{M}(\vec{r}),\hat{\rho}]]}{\norm{\vec{r}-\vec{r}'}}    
\label{diosipenrose}
\end{equation}
with the gravitational constant $G$, the total mass density operator $\hat{M}$, and the Euclidean norm $\norm{\cdot}$. This operator is given by \cite{DonadiPCDLB2020}
\begin{equation}
\hat{M}(\vec{r})=\sum_{i=1}^{N}\hat{\mu}_i(\vec{r})
\end{equation}
with the mass-density operator $\hat{\mu}_i$ for the $i$-th particle. The Di{\'o}si-Penrose model constitutes a third type of collapse models in addition to GRW theory and CSL model.

A difficulty arises from the fact that the standard definition 
\begin{equation}
\hat{\mu}_i(\vec{r})=m_i\delta(\vec{r}-\hat{\vec{r}}_i)    
\end{equation}
with the mass $m_i$ of and the position operator $\hat{\vec{r}}_i$ for the $i$-th particle leads to divergences \cite{DonadiPCDLB2020}. To solve this problem, one has to smear out the mass density. This can be done in various ways \cite{BahramiSB2014}. Proposals from the literature include \cite{Diosi1989}
\begin{equation}
\hat{\mu}_i(\vec{r})=\frac{3m_i}{4\pi R_0^2}\Theta(R_0 - \norm{\vec{r}-\hat{\vec{r}}}), 
\end{equation}
with a length scale $R_0$ and the Heaviside function $\Theta$, 
as well as \cite{GhirardiGR1990} 
\begin{align}
\hat{\mu}_i(\vec{r})=\frac{m_i}{(2\pi R_0^2)^{\frac{3}{2}}}\exp\!\bigg(-\frac{(\vec{r}-\hat{\vec{r}}_i)^2}{2R_0^2}\bigg).
\end{align}

We are only interested in the non-Hamiltonian term on the right-hand side of \cref{diosipenrose}. Moreover, we restrict ourselves to a single particle with mass density operator $\hat{\mu}$. As shown in the supplementary material of Ref.\ \cite{DonadiPCDLB2020}, the non-Hamiltonian term in \cref{diosipenrose} can be written as
\begin{equation}
\INT{}{}{^3p'}\tilde{\Gamma}(\vec{p}')\Big(e^{\frac{\ii}{\hbar}\vec{p}'\cdot\hat{\vec{r}}}\hat{\rho}\, e^{-\frac{\ii}{\hbar}\vec{p}'\cdot\hat{\vec{r}}} - \hat{\rho}\Big)   
\label{diosipenrose2}
\end{equation}
with the function
\begin{equation}
\tilde{\Gamma}(\vec{p}) =\frac{4 G}{\pi \hbar^2}\frac{|\tilde{\mu}(\vec{p})|^2}{\norm{\vec{p}}^2}
\label{gammatilde}
\end{equation}
depending on the Fourier-transformed mass density $\tilde{\mu}$.  
The matrix element of \cref{diosipenrose2} is given by \cite{DonadiPCDLB2020}
\begin{equation}
\INT{}{}{^3p'}\tilde{\Gamma}(\vec{p}')\Big(e^{\frac{\ii}{\hbar}\vec{p}'\cdot(\vec{r}-\vec{r}')}-1\Big)\braket{\vec{r}|\hat{\rho}|\vec{r}'}.    
\end{equation}
Hence, the Weyl symbol of the expression \eqref{diosipenrose2} is given by
\begin{align}
&\frac{1}{(2\pi\hbar)^3}\INT{}{}{^3r'}\INT{}{}{^3p'}\tilde{\Gamma}(\vec{p}')\Big(e^{-\frac{\ii}{\hbar}\vec{p}'\cdot\vec{r}'}-1\Big)\notag\\
&\bigg\langle\vec{r}-\frac{\vec{r}'}{2}
\bigg\lvert\hat{\rho}\bigg\rvert\vec{r} 
+ \frac{\vec{r}'}{2}\bigg\rangle e^{\frac{\ii}{\hbar}\vec{p}\cdot\vec{r}'}\notag\\
&=\frac{1}{(2\pi\hbar)^3}\INT{}{}{^3r'}\INT{}{}{^3p'}\tilde{\Gamma}(\vec{p}')\bigg\langle\vec{r}
-\frac{\vec{r}'}{2}\bigg\lvert\hat{\rho}\bigg\rvert\vec{r} 
+ \frac{\vec{r}'}{2}\bigg\rangle e^{\frac{\ii}{\hbar}(\vec{p}-\vec{p}')\cdot\vec{r}'}  \notag\\
&\quad\, - W(\vec{r},\vec{p})\notag\\
&=\INT{}{}{^3p'}\tilde{\Gamma}(\vec{p}')(W(\vec{r},\vec{p}-\vec{p}')  - W(\vec{r},\vec{p})).
\label{diosipenrose3}
\end{align}
Using \cref{diosipenrose3}, the \textit{Di{\'o}si-Penrose master equation for Wigner functions} is given by
\begin{align}
\partial_t W(\vec{r},\vec{p}) &= \{H(\vec{r},\vec{p}), W(\vec{r},\vec{p})\}_\star \label{diosipenrose4}\\
&\quad\, + \INT{}{}{^3p'}\tilde{\Gamma}(\vec{p}')(W(\vec{r},\vec{p}-\vec{p}') - W(\vec{r},\vec{p})). \notag
\end{align}
When comparing \cref{diosipenrose4} with \cref{grwwigner}, it is easily seen that the general structure is the same, in particular if we assume that the mass-density distribution is Gaussian. However, the expression that the Wigner function is convoluted with differs due to the presence of the factor $1/(\norm{\vec{p}}')^2$ in \cref{gammatilde}. Equation \eqref{diosipenrose4} is an important result, since it has recently been suggested that Wigner functions could be helpful for experimental tests of quantum gravity \cite{HowlVNCRI2021}.

\section{\label{dgm}Dissipative GRW model}
Finally, we turn to a more recent extension of the GRW theory, the \textit{dissipative GRW model} proposed by \citet{SmirneVB2014}. As discussed in \cref{grwtheory}, the standard spontaneous collapse models have the problem that the energy increases continuously. This increase is not measurable on typical observational timescales, but leads to a divergence of the temperature for $t\to\infty$. To avoid this problem, the dissipative GRW model includes dissipation to ensure that the system thermalizes to a finite temperature. A similar extension has been developed for the CSL model by \citet{SmirneB2015} and for the Di{\'o}si-Penrose model by \citet{BahramiSB2014}.

Since the derivation of the Wigner representation is more involved for this theory, we restrict the discussion to a single particle in one dimension to improve readability (the extension is straightforward). The dissipative GRW model is given by the master equation \cite{SmirneVB2014}
\begin{widetext}
\begin{equation}
\partial_t \hat{\rho} = \frac{\ii}{\hbar}[\hat{H},\hat{\rho}] - \lambda(\hat{\rho} - \hat{T}_\mathrm{d}(\hat{\rho})),
\label{dissipativegrw}
\end{equation}
which looks exactly like \cref{grw}. However, the standard collapse operator $\hat{T}$ is replaced by the dissipative collapse operator
\begin{equation}
\hat{T}_\mathrm{d}(\hat{\rho}) = \frac{r_{\mathrm{c}}(1+k)}{\sqrt{\pi}\hbar}\INT{}{}{p'}e^{\frac{\ii}{\hbar}p'\hat{x}}e^{-\frac{r_{\mathrm{c}}^2}{2\hbar^2}((1+k)p'+2k\hat{p})^2}\hat{\rho}\, e^{-\frac{r_{\mathrm{c}}^2}{2\hbar^2}((1+k)p'+2k\hat{p})^2}e^{-\frac{\ii}{\hbar}p'\hat{x}}  
\label{dissipationoperator}
\end{equation}
with a localization length scale $r_{\mathrm{c}} = 1/\sqrt{\alpha}$ and a temperature parameter $k$. The operator \eqref{dissipationoperator} reduces to the standard collapse operator \eqref{standardcollapseoperator} (and the dissipative GRW model \eqref{dissipativegrw} to the ordinary GRW model \eqref{grw}) for $k \to 0$, which corresponds to the high-temperature limit. 

In the momentum representation, \cref{dissipationoperator} gives
\begin{align}
\braket{p|\hat{T}_\mathrm{d}|q} &= \bigg\langle p\bigg\lvert \frac{r_{\mathrm{c}}(1+k)}{\sqrt{\pi}\hbar}\INT{}{}{p'}e^{\frac{\ii}{\hbar}p'\hat{x}}e^{-\frac{r_{\mathrm{c}}^2}{2\hbar^2}((1+k)p'+2k\hat{p})^2}\hat{\rho}\,
e^{-\frac{r_{\mathrm{c}}^2}{2\hbar^2}((1+k)p'+2k\hat{p})^2}e^{-\frac{\ii}{\hbar}p'\hat{x}}\bigg\rvert q\bigg\rangle \notag\\
&= \frac{r_{\mathrm{c}}(1+k)}{\sqrt{\pi}\hbar}\INT{}{}{p'} \bigg\langle p - p'\bigg\lvert e^{-\frac{r_{\mathrm{c}}^2}{2\hbar^2}((1+k)p'+2k\hat{p})^2}\hat{\rho}\, 
e^{-\frac{r_{\mathrm{c}}^2}{2\hbar^2}((1+k)p'+2k\hat{p})^2}\bigg\rvert q - p'\bigg\rangle\\
&= \frac{r_{\mathrm{c}}(1+k)}{\sqrt{\pi}\hbar}\INT{}{}{p'} e^{-\frac{r_{\mathrm{c}}^2}{2\hbar^2}((1+k)p'+2k(p-p'))^2}e^{-\frac{r_{\mathrm{c}}^2}{2\hbar^2}((1+k)p'+2k(q-p'))^2} 
\big\langle p - p'\big\lvert\hat{\rho}\big\rvert q - p'\big\rangle \notag\\
&= \frac{r_{\mathrm{c}}(1+k)}{\sqrt{\pi}\hbar}\INT{}{}{p'} e^{-\frac{r_{\mathrm{c}}^2}{2\hbar^2}(4k^2p^2 + 4kp p' - 4 k^2 p p' + 2 (p')^2 - 4 k (p')^2 + 2 k^2 (p')^2 + 4 k p' q - 4 k^2 p' q + 4 k^2 q^2)}\big\langle p - p'\big\lvert\hat{\rho}\big\rvert q - p'\big\rangle ,\notag
\end{align}
where $\ket{p}$ and $\ket{q}$ are momentum eigenstates with eigenvalues $p$ and $q$, respectively. We now use the fact that the Wigner function \eqref{Wigner} can also be written as \cite{teVrugtW2019c}
\begin{equation}
W(x,p) = \frac{1}{2\pi\hbar}\INT{}{}{p''}\bigg\langle{p-\frac{1}{2}p''
\bigg\lvert\hat{\rho}\bigg\rvert
p+\frac{1}{2}p''\bigg\rangle e^{-\frac{\ii}{\hbar}x p''}}.
\label{Wigner2}%
\end{equation}
Thus, the Weyl symbol of $\hat{T}_\mathrm{d}$ is given by
\begin{align}
&T_\mathrm{d}(W(x,p)) \notag\\
&=  \frac{r_{\mathrm{c}}(1+k)}{\sqrt{\pi}\hbar}\frac{1}{2\pi\hbar}\INT{}{}{p'}\INT{}{}{p''} e^{-\frac{r_{\mathrm{c}}^2}{2\hbar^2}(8k^2p^2 + 8 k p p' - 8 k^2 p p' + 2 (p')^2 - 4 k (p')^2 + 2 k^2 (p')^2 + 2 k^2 (p'')^2)}e^{-\frac{\ii}{\hbar}x p''}\bigg\langle p - p' - \frac{p''}{2}\bigg\lvert\hat{\rho}\bigg\rvert p- p'+\frac{p''}{2}\bigg\rangle \notag\\
&=  \frac{r_{\mathrm{c}}(1+k)}{\sqrt{\pi}\hbar}\frac{1}{2\pi\hbar}\frac{1}{\sqrt{2 \pi k^2 r_{\mathrm{c}}^2}} 
\INT{}{}{p'}\INT{}{}{p''}\INT{}{}{x'} \label{dissipativeweylsymbol}\\
&\quad\; e^{-\frac{r_{\mathrm{c}}^2}{2\hbar^2}(8k^2p^2 + 8 k p p' - 8 k^2 p p' + 2 (p')^2 - 4 k (p')^2 + 2 k^2 (p')^2)}e^{-\frac{(x')^2}{2k^2 r_{\mathrm{c}}^2}}e^{-\frac{\ii}{\hbar}(x-x') p''}\bigg\langle p - p' - \frac{p''}{2}\bigg\lvert\hat{\rho}\bigg\rvert p- p'+\frac{p''}{2}\bigg\rangle \notag\\
&=  \frac{r_{\mathrm{c}}(1+k)}{\sqrt{\pi}\hbar}\frac{1}{\sqrt{2\pi k^2 r_{\mathrm{c}}^2}}\INT{}{}{p'}\INT{}{}{x'} e^{-\frac{r_{\mathrm{c}}^2}{2\hbar^2}(8k^2p^2 + 8 k p p' - 8 k^2 p p' + 2 (p')^2 - 4 k (p')^2 + 2 k^2 (p')^2)}e^{-\frac{(x')^2}{2k^2 r_{\mathrm{c}}^2}}W(x-x',p-p').\notag
\end{align}
Hence, the \textit{dissipative GRW master equation for Wigner functions} reads
\begin{equation}
\partial_t W(x,p) = \{H(x,p),W(x,p)\}_\star - \lambda (W(x,p) - T_\mathrm{d}(W(x,p)))   \label{dissipativegrwmodelwigner} 
\end{equation}
with $T_\mathrm{d}$ given by \cref{dissipativeweylsymbol}.

Since \cref{dissipativegrwmodelwigner} with the Weyl symbol \eqref{dissipativeweylsymbol} is extremely complicated, we now derive a simplified form. First, we note that \cref{dissipativeweylsymbol} can be written as 
\begin{equation}
\begin{split}
&T_\mathrm{d}(W(x,p))\\
&=\frac{r_{\mathrm{c}}(1+k)}{\sqrt{\pi}\hbar}\frac{1}{2\pi\hbar}\INT{}{}{p'}\INT{}{}{p''} e^{-\frac{r_{\mathrm{c}}^2}{2\hbar^2}(8k^2p^2 + 8 k p p' - 8 k^2 p p' + 2 (p')^2 - 4 k (p')^2 + 2 k^2 (p')^2)} \\
&\quad\;\, e^{-\frac{k^2 r_{\mathrm{c}}^2 (p'')^2}{\hbar^2}}e^{-\frac{\ii}{\hbar}xp''}
\bigg\langle p - p' - \frac{p''}{2}\bigg\lvert\hat{\rho}\bigg\rvert p- p'+\frac{p''}{2}\bigg\rangle.
\end{split}
\label{dissipativeweylsymbola}%
\end{equation}
(Equation \eqref{dissipativeweylsymbola} is simply the first line of \cref{dissipativeweylsymbol}.) Motivated by the fact that the standard (nondissipative) GRW model is recovered for $k=0$, we now Taylor expand \cref{dissipativeweylsymbola} to first order in $k$. This gives
\begin{equation}
\begin{split}
&T_\mathrm{d}(W(x,p))\\
&\approx \frac{r_{\mathrm{c}}(1+k)}{\sqrt{\pi}\hbar}\frac{1}{2\pi\hbar}\INT{}{}{p'}\INT{}{}{p''}\bigg(1-\frac{r_{\mathrm{c}}^2}{2\hbar^2}(8 k p p' - 4 k (p')^2)\bigg)e^{-\frac{r_{\mathrm{c}}^2(p')^2}{\hbar^2}}e^{-\frac{\ii}{\hbar}xp''}\bigg\langle p - p' - \frac{p''}{2}\bigg\lvert\hat{\rho}\bigg\rvert p- p'+\frac{p''}{2}\bigg\rangle\\
&=\frac{r_{\mathrm{c}}(1+k)}{\sqrt{\pi}\hbar}\INT{}{}{p'}\bigg(1-\frac{r_{\mathrm{c}}^2}{2\hbar^2}(8 k p p' - 4 k (p')^2)\bigg)e^{-\frac{r_{\mathrm{c}}^2(p')^2}{\hbar^2}}W(x,p-p').
\end{split}
\label{dissipativeweylsymbolb}
\end{equation}
The result \eqref{dissipativeweylsymbolb} is already reminiscent of \cref{ttransformation} from the standard GRW model. In analogy to the treatment in \cref{grwtheory}, we now Taylor expand $W(x,p-p')$ around $p=p'$ in \cref{dissipativeweylsymbolb} and find
\begin{equation}
\begin{split}
&T_\mathrm{d}(W(x,p))\\
&=\frac{r_{\mathrm{c}}(1+k)}{\sqrt{\pi}\hbar}\INT{}{}{p'}\bigg(1-\frac{r_{\mathrm{c}}^2}{2\hbar^2}(8 k p p' - 4 k (p')^2)\bigg)e^{-\frac{r_{\mathrm{c}}^2(p')^2}{\hbar^2}}\Big(W(x,p)-\partial_p W(x,p)p' + \frac{1}{2}\partial_p^2W(x,p)(p')^2\Big)\\
&\approx (1+2k)W(x,p) +2 k p\partial_p W(x,p)+ \frac{\hbar^2}{4r_{\mathrm{c}}^2}\partial_p^2 W(x,p),
\end{split}
\label{dissipativeweylsymbolc}
\end{equation}
where we have dropped terms of order $k^2$ and products of $k$ and $\hbar^2$ in the last step. After inserting \cref{dissipativeweylsymbolc} into \cref{dissipativegrwmodelwigner}, we finally obtain the Fokker-Planck equation
\begin{equation}
\partial_t W(x,p) = \{H(x,p),W(x,p)\}_\star + \gamma \partial_p (p W(x,p)) + \gamma m k_{\mathrm{B}} T_\mathrm{n} \partial_p^2 W(x,p) 
\label{grwkramers}
\end{equation}
with damping parameter $\gamma = 2\lambda k$, Boltzmann constant $k_{\mathrm{B}}$, and noise temperature $T_\mathrm{n} = \hbar^2 /(8 m k k_{\mathrm{B}} r_{\mathrm{c}}^2)$. This is precisely the temperature a dissipative GRW system thermalizes to \cite{TorovsB2018}, which confirms the consistency of our approach. If we ignore the fact that \cref{grwkramers} contains a Moyal rather than a Poisson bracket, it is formally identical to the \textit{Kramers equation} \cite{Risken1996}
\begin{equation}
\partial_t W(x,p) = \Big(- \frac{p}{m}\partial_x + (\partial_x U(x)) \partial_p + \partial_p\gamma p + \gamma m k_{\mathrm{B}} T_\mathrm{b} \partial_p^2 \Big)W(x,p),
\label{kramers}
\end{equation}
which describes the dynamics of an underdamped Brownian particle in contact with a heat bath $T_\mathrm{b}$. In \cref{grwkramers}, however, the damping and diffusion terms arise not from the interaction with a solvent, but from the fundamental equations of motion of (GRW-type) quantum mechanics, implying that they are also present in a closed system.
\end{widetext}

\section{\label{irre}Irreversibility and approach to equilibrium}
David Albert \cite{Albert1994,Albert1994b,Albert2000} has suggested that the GRW theory might be useful in explaining the approach to thermodynamic equilibrium. A typical explanation for the approach to thermodynamic equilibrium runs as follows: Macroscopic states of a system are associated with a large number of possible microscopic states. The equilibrium state is the macrostate with the largest entropy, i.e., the state with the largest number of possible microscopic realizations. Consequently, most of the microstates associated with an initial macrostate will evolve towards an equilibrium state, and since we do not know the exact microstate of the system, we have to conclude that it is considerably more likely that the system will evolve towards equilibrium rather than away from it.

This explanation, Albert argues, is unsatisfactory because it involves an appeal to the limitations of observers (ignorance about the precise initial microstate) in explaining what appears to be an observer-independent phenomenon, namely the approach to equilibrium. GRW theory provides a possible solution by introducing objective probabilities associated with the spontaneous collapse of the wavefunction. If the system is initially in an \ZT{abnormal} microstate that would lead to an evolution away from equilibrium -- which is unlikely, but not impossible -- then a GRW collapse will in most cases lead to a \ZT{normal} state from which the system equilibrates. According to Albert, this happens since every microscopic neighborhood of the abnormal microstate in phase space contains significantly more normal than abnormal microstates, such that thermodynamic abnormality is unstable under perturbations. Some discussions of Albert's proposal can be found in Refs.\ \cite{HemmoS2001,HemmoS2003,HemmoS2005}. See Ref.\ \cite{North2011} for a review.

Albert's approach belongs to the so-called \textit{Boltzmann approach} to statistical mechanics \cite{HemmoS2003}. In this framework, one is interested in a single system that has a certain macrostate (characterized by its macroscopic properties). This macrostate depends on the microscopic state of the system. Various microstates can correspond to the same macrostate. The equilibrium state is then the macrostate that the largest number of microstates correspond to. An alternative framework, which is more widely used in practical applications, is the \textit{Gibbs approach}, where one studies ensembles (hypothetical sets of infinitely many copies of a system with different initial conditions). \ZT{Approach to equilibrium} then means that the probability distribution over an ensemble of systems converges towards the equilibrium (in the simplest case: Maxwell-Boltzmann) distribution. A more detailed discussion and comparison of both approaches can be found in Ref.\ \cite{Frigg2008}. 

Albert's proposal is conceptually appealing for at least three reasons \cite{HemmoS2003}: First, it allows to solve two problems -- the quantum measurement problem and the problem of thermodynamic irreversibility -- at once. Second, it unifies quantum-mechanical probabilities with the probabilities used in statistical mechanics. Third, it allows to explain thermodynamic irreversibility without requiring assumptions about a probability distribution over initial conditions. However, it is only based on a qualitative argument, a quantitative test is lacking at present. Such a test would be required to see how far GRW collapses can actually get us on the road to equilibrium \cite{North2011}.

A simple argument for Albert's proposal can be developed based on \cref{grwfokkermany}. As is well known, the \textit{Boltzmann equation} \cite{Boltzmann1872} provides a useful model for the approach to thermodynamic equilibrium. The one-particle distribution function $f(\vec{p})$, giving the probability that a particle has momentum $\vec{p}$, approaches the Maxwell-Boltzmann (equilibrium) distribution due to the presence of collisions. This is a consequence of the fact that, in the derivation of the Boltzmann equation, one assumes that velocities are uncorrelated, i.e., that the two-particle distribution satisfies $f_2(\vec{p}_1,\vec{p}_2) = f(\vec{p}_1)f(\vec{p}_2)$ (molecular chaos), before a collision \cite{BrownUM2009}. If one instead assumes that velocities are uncorrelated \textit{after} a collision, we would have found that the entropy always decreases. 

Thus, if \cref{grwfokkermany} should explain the approach to thermodynamic equilibrium, it should enforce $f_2(\vec{p}_1,\vec{p}_2) = f(\vec{p}_1)f(\vec{p}_2)$. There is a good reason to believe that this is the case: The distributions $f$ and $f_2$ can be obtained by integrating the Wigner function\footnote{Since $f$ has been interpreted here as a probability distribution (which is not generally possible), we have to assume that a classical limit can be and has been taken in such a way that $W$ reduces to a classical probability distribution. Note, however, that a Wigner-Boltzmann equation also exists \cite{SelsB2013}. It is a transport equation for the Wigner function used, e.g., in semiconductor physics \cite{NedjalkovKSRF2004}.} over all positional and over all but one (for $f$) or two (for $f_2$) momentum degrees of freedom \cite{CancellieriBJ2007}. Ignoring the reversible Hamiltonian part of the dynamics, we find that the difference between a factorized and a nonfactorized distribution evolves as
\begin{equation}
\begin{split}
&\partial_t (f_2(\vec{p}_1,\vec{p}_2) - f(\vec{p}_1)f(\vec{p}_2))\\ 
&=D_p(\Nabla_{\vec{p}_1}^2
+ \Nabla_{\vec{p}_2}^2)(f_2(\vec{p}_1,\vec{p}_2) - f(\vec{p}_1)f(\vec{p}_2)).
\end{split}
\label{difference}
\end{equation}
Obviously, $f_2(\vec{p}_1,\vec{p}_2) = f(\vec{p}_1)f(\vec{p}_2)$ is a fixed point of \cref{difference}. From the general properties of the diffusion equation (which leads to a homogeneous distribution), we can infer that the fixed point is stable. Consequently, the GRW dynamics indeed tends to destroy correlations in momentum space and thereby leads to molecular chaos.

There is also a different aspect that has to be taken into account: When comparing \cref{grwfokkermany} to the Kramers equation \eqref{kramers} we can note that \cref{grwfokkermany} has no friction term. However, the friction term in \cref{kramers} is known to be essential for the approach to an equilibrium state (corresponding, in the classical case, to the Maxwell-Boltzmann distribution), and is intimately connected to the fluctuations described by the diffusion term in \cref{kramers} via a fluctuation-dissipation theorem. The fluctuations in \cref{grwfokkermany}, in contrast, are not connected to any form of dissipation. Thus, if \cref{grwfokkermany} induces equilibration, it will be of a different form than equilibration processes resulting from friction.

This brings us back to the distinction between Boltzmannian and Gibbsian approaches to equilibrium introduced above. If $W$ can be interpreted as a probability distribution (i.e., in the classical case), the Kramers equation \eqref{kramers} describes an approach towards the Maxwell-Boltzmann distribution, which is the equilibrium distribution of Gibbsian statistical mechanics. For this, the friction term plays an essential role. In \cref{grwfokkermany}, in contrast, a friction term is not present, such that we cannot expect that $W$ converges to a Maxwell-Boltzmann distribution. This, however, does not exclude that it leads (with high probability) to a convergence towards the macrostate with the largest phase-space volume, i.e., to Boltzmannian equilibrium.

Up to now, we have only considered the original GRW model by \citet{GhirardiRW1986}. This is the theory that Albert's original suggestion is based on.\footnote{Albert's theory was formulated in 1994 \cite{Albert1994,Albert1994b}, whereas the dissipative GRW model was developed in 2014 \cite{SmirneVB2014}.} The drawbacks of the standard GRW model in explaining equilibration can, however, be avoided by considering the dissipative GRW model introduced in \cref{dgm}. As shown there, the Wigner function approximately follows the dynamic equation \eqref{grwkramers} in this theory. Since \cref{grwkramers} \textit{does} contain a damping term related to the diffusion term in the standard way (actually, \cref{grwkramers} reduces to \cref{kramers} for $\hbar \to 0$ if we set $T_\mathrm{n} = T_\mathrm{b}$), the dissipative GRW model should be perfectly capable of describing the approach to thermodynamic equilibrium. This suggests the following modification of Albert's proposal: The approach to thermodynamic equilibrium might be explained by spontaneous \textit{dissipative} GRW collapses.

This would have interesting consequences: The Kramers equation \eqref{kramers}, of which \cref{grwkramers} is an extension, describes a system coupled to a heat bath at fixed temperature (rather than an isolated system with fixed energy) that should be described in the canonical ensemble. Since spontaneous GRW-type collapses occur also in isolated systems, this would imply that the equilibrium distribution of an \textit{isolated} system is given by the canonical rather than by the microcanonical distribution \cite{SmirneB2015,BahramiSB2014}. This is a very significant difference to \ZT{standard} statistical mechanics, where isolated systems are described by a microcanonical distribution, which corresponds to the energy-conserving case. While the different ensembles agree in the thermodynamic limit, they can differ significantly for systems with a small number of particles \cite{delasHerasBFS2016}. A different result would be obtained in an energy-conserving spontaneous collapse model that might be derived in the future. (Note, however, that Albert's original discussion \cite{Albert1994,Albert1994b} is set up in a Boltzmannian framework that is not based on ensembles at all.) In addition, the reduction of the spontaneous heating by dissipation has implications for tests of spontaneous collapse models that rely on this effect, for example those based on measurements of neutron star heating \cite{TilloyS2019}.

The idea that the dissipative GRW model \eqref{grwkramers} is more appropriate as an explanation for thermodynamic irreversibility than the original GRW model is, however, solely based on the fact that it is formally analogous to \cref{kramers}, which is known to lead to equilibration. Physically, the dissipative GRW model essentially assumes that all particles in the universe are coupled to a universal heat bath with a certain temperature $T_\mathrm{n}$. (The nondissipative GRW model is the special case $T_\mathrm{n} \to \infty$.) Consequently, in the dissipative GRW model, systems will (in the long run) approach a canonical equilibrium state with temperature $T_\mathrm{n}$. This is irreversibility, but not of the type known from classical thermodynamics. There, a typical manifestation of equilibration is that if two systems with initially different temperatures $T_1$ and $T_2$ are brought in thermal contact, heat will flow from the hotter to the colder system until they have reached equal temperatures \cite{teVrugt2021}. This final temperature will depend on the initial temperatures and other properties of the systems. It is certainly not a prediction of classical thermodynamics that all isolated systems converge to the \textit{same} equilibrium state characterized by a universal temperature $T_\mathrm{n}$.

This, of course, does not necessarily imply that the dissipative GRW model is incorrect. It merely implies that it makes a different empirical prediction than standard thermodynamics, namely that there is a universal temperature that all systems, irrespective of their initial temperature and energy, will equilibrate to. This equilibration will take place on a very long timescale, such that this prediction does not contradict the observation of \ZT{normal} equilibration in the form of heat exchange on ordinary timescales. However, it was this equilibration that we were seeking an explanation for. In this regard, the dissipative GRW model will play the same role (if any) as the standard one, namely that of erasing correlations by providing dissipation.

These conclusions are not affected in a significant way by working (as Albert does) in a Boltzmannian rather than a Gibbsian framework. As discussed by \citet{teVrugt2020}, the distinction between Gibbsian and Boltzmannian statistical mechanics can be linked to the distinction between deterministic and stochastic forms of dynamical density functional theory (DDFT) \cite{teVrugtLW2020}. DDFT is a theory for the nonequilibrium evolution of the one-body density in classical fluids. Deterministic DDFT describes ensembles, whereas stochastic DDFT a (possibly spatially averaged) single system \cite{ArcherR2004}. We can thus get an idea of the Boltzmannian equilibration predicted by dissipative GRW models by considering the predictions of stochastic DDFT, namely an approach to the minimum state of the free energy functional (which has a different definition compared to deterministic DDFT \cite{teVrugtLW2020}). Hence, according to dissipative GRW theory, the system will (in equilibrium) fluctuate around the minimum of a free energy functional\footnote{In Gibbsian statistical mechanics, it is well known that equilibrium systems coupled to a heat bath minimize a free energy functional. For the Boltzmannian case, this deserves some comment: As shown by \citet{Dean1996}, a system of Langevin equations for the position of classical particles can be exactly rewritten as a Langevin equation for the microscopic one-body density. The latter Langevin equation can be written in terms of a (free energy) functional depending on the one-body density. Then, the corresponding Fokker-Planck equation has a stationary solution if one inserts the one-body density that leads to the minimum of this functional. As discussed in \cref{numeric}, the classical limit of dissipative GRW theory also leads to Langevin equations, with the difference to Dean's treatment being that we have to take also the momentum density into account. Dean's formalism has been extended to underdamped Langevin equations (for which the momentum density is relevant) by \citet{NakamuraY2009}.} with temperature $T_\mathrm{n}$.

When it comes to solving the fundamental problems of explaining thermodynamic irreversibility, typical arguments against coarse-graining -- namely that the equilibrium state of equilibrium statistical mechanics is reproduced only approximately \cite{RidderbosR1998} and on certain timescales -- also apply to GRW-based explanations. Nevertheless, Albert's approach, if successful, would still explain why anti-thermodynamic behavior is never observed in macroscopic systems. For microscopic systems, it has recently been demonstrated experimentally that heat can spontaneously flow from a cold to a hot system for initially correlated spins \cite{MicadeiEtAl2019}. This is not in contradiction to the GRW-based approach, since GRW collapses are (by construction) extremely rare for small quantum systems and thus allow for such a behavior.

\section{\label{deco}Connection to decoherence}
Spontaneous collapse models like the GRW theory have a very close connection to the theory of quantum decoherence \cite{Vacchini2007}, which is reviewed in Refs.\ \cite{Zurek2003,JacquodP2009,Schlosshauer2005,NarnhoferW2014}. In fact, some authors refer to GRW-type collapses as \ZT{intrinsic decoherence} \cite{Stamp2012}. Moreover, there has been a significant amount of research on explaining the emergence of irreversibility based on the interaction of quantum systems with their environment \cite{LindenPSW2009}, in particular decoherence \cite{JacquodP2009,NarnhoferW2014,ZurekP1994}. Consequently, it is of interest to relate such approaches to Albert's proposal discussed here. The purpose of this section (which is a slight digression) is thus to present some other approaches from the literature that Albert's theory, on which we focus here, is related to. This allows to see our results in a more general context.

In contrast to the GRW theory, decoherence does by itself neither provide a solution to the measurement problem nor produce stochasticity. However, it is an important ingredient in various interpretations of quantum mechanics \cite{Wallace2012}. This has been exploited in an explanation of the approach to equilibrium proposed by Hemmo and Shenker \cite{HemmoS2001,HemmoS2003,HemmoS2005}. Like Albert's idea, it is based on explaining probabilities in statistical mechanics based on quantum-mechanical probabilities. However, they take as a starting point not GRW theory, but no-collapse interpretations (such as modal interpretations or the Everett interpretation), in which a collapse of the wavefunction does not occur. Since the quantum dynamics of closed systems is isolated in the absence of collapses, stochasticity then requires external interventions. Thus, the proposal discussed in this section belongs to the tradition of \textit{interventionism} \cite{RidderbosR1998,Ridderbos2002,Blatt1959}, which explains the approach to equilibrium based on external perturbations.

We discuss the approach following Ref.\ \cite{HemmoS2003}. In decoherence theory, one typically assumes that the initial state $\ket{\Psi}$ of a quantum system and its environment can be written in the form
\begin{equation}
\ket{\Psi}(0)=\ket{\psi}\otimes\ket{E} 
\label{psio}
\end{equation}
with the state of the system $\ket{\psi}$, the state of the environment $\ket{E}$, and the tensor product $\otimes$. Moreover, one assumes that the interaction Hamiltonian describing the interaction of the system with its environment commutes with a certain observable, the so-called \textit{pointer variable}. A typical pointer variable is the position. Let $\{\ket{\psi}_i\}$ be a basis of the Hilbert space of the quantum system that consists of eigenstates of the pointer variable. The state of the system at time $t$ can then be written as
\begin{equation}
\ket{\Psi}(t)=\sum_{i}\nu_i(t)\ket{\psi}_i\otimes\ket{E}_i     
\end{equation}
with the relative states $\ket{E}_i$ and the amplitudes $\nu_i(t)$. Due to the coupling with the environment, the relative states will satisfy $\braket{E_i|E_j} \approx \delta_{ij}$ with the Kronecker delta $\delta_{ij}$ after an extremely short time. Then, the reduced density operator (obtained by tracing over environmental degrees of freedom) of the system takes the form
\begin{equation}
\hat{\rho}(t) = \sum_{i}|\nu_i(t)|^2 \ket{\psi}_i\bra{\psi}_i. 
\label{reducedstate}%
\end{equation}
For a harmonic oscillator weakly coupled to an environment in equilibrium, the states $\ket{\psi}_i$ correspond to narrowly peaked Gaussians, such that the decohered system follows quasi-classical trajectories \cite{ZurekHP1993}.

This result is, by itself, no solution of the quantum measurement problem since it leads to a superposition of different classical time evolutions. However, in no-collapse interpretations of quantum mechanics, such as modal interpretations \cite{BacciagaluppiD1999}, it is possible to interpret the reduced state \eqref{reducedstate} as a probability distribution over the different quasi-classical states $\{\ket{\psi}_i\}$. During the time evolution, there will then be stochastic transitions between such states. These stochastic transitions can then play the same role in the approach to equilibrium as the stochastic GRW collapses in Albert's theory.

What is interesting about this approach is that the Wigner function of a decohering system coupled to a thermal environment can, in the simplest case, be shown to have the form \cite{HemmoS2003,ZurekP1994,ZurekP1995,CaldeiraL1983,Halliwell2007,Tegmark1993}
\begin{equation}
\begin{split}
\partial_t W(x,p) &= \{H(x,p),W(x,p)\}_\star\\ 
&\quad\, +(\partial_p\gamma p+ \gamma m k_{\mathrm{B}} T_\mathrm{b} \partial_p^2) W(x,p),
\end{split}
\label{momentumdiffusion2}
\end{equation}
which, for $\hbar \to 0$, reduces to the Kramers equation \eqref{kramers}. Thus, in this approach, the probability distribution also approaches a Maxwell-Boltzmann form in the classical case. Consequently, it is, like the dissipative GRW model, close in form to standard theories of equilibration. However, for these models, the temperature the system equilibrates to is that of the environment (rather than that of a universal heat bath).

Determining which of these options is the correct one (if any) is beyond the scope of this work. In particular, it would obviously require an experimental test \cite{VinanteCBCVFMMU2020,DonadiPCDLB2020,ZhengEtAl2020} of the (dissipative) GRW model. Moreover, it should be noted that there are important conceptual differences between these approaches -- if we explain thermodynamic irreversibility by decoherence, then there can be no irreversibility in an isolated system. In contrast, such a behavior is to be expected from the GRW perspective, which would thus (in a sense) imply that the standard models of stochastic dynamics (such as the Fokker-Planck equation or the Langevin equation) are actually more fundamental than the ordinary Hamiltonian dynamics which they are usually thought to be an approximation to. 

A further recent approach to the problem of explaining the emergence of statistical mechanics that is interesting in this context was proposed by Drossel \cite{Drossel2020,Drossel2017,Drossel2015,DrosselE2018}. She argues that the practice of statistical mechanics, including (but not limited to) the presence of irreversibility, is incompatible with the universal validity of deterministic unitary quantum mechanics, and has to be explained by the presence of fundamental stochasticity. Her view can be classified as an intermediate position between that of Albert and that of Hemmo and Shenker. Like Albert (and unlike Hemmo and Shenker), she sees the origin of irreversibility in a stochastic quantum dynamics that violates the standard Schr\"odinger time evolution. However, like Hemmo and Shenker (and unlike Albert), she attributes this stochasticity to interactions of a quantum system with the environment. Finally, \citet{Wallace2016} has argued, within an Everettian framework, that the probabilities of statistical mechanics can be understood as arising from quantum-mechanical probabilities based on the fact that the classical limit of quantum mechanics is approximately isomorphic to a theory of classical probability distributions (as is evident from the Wigner function picture).

\section{\label{numeric}Numerical experiments}
\begin{figure*}[tbhp]
\centering\includegraphics[width=\linewidth]{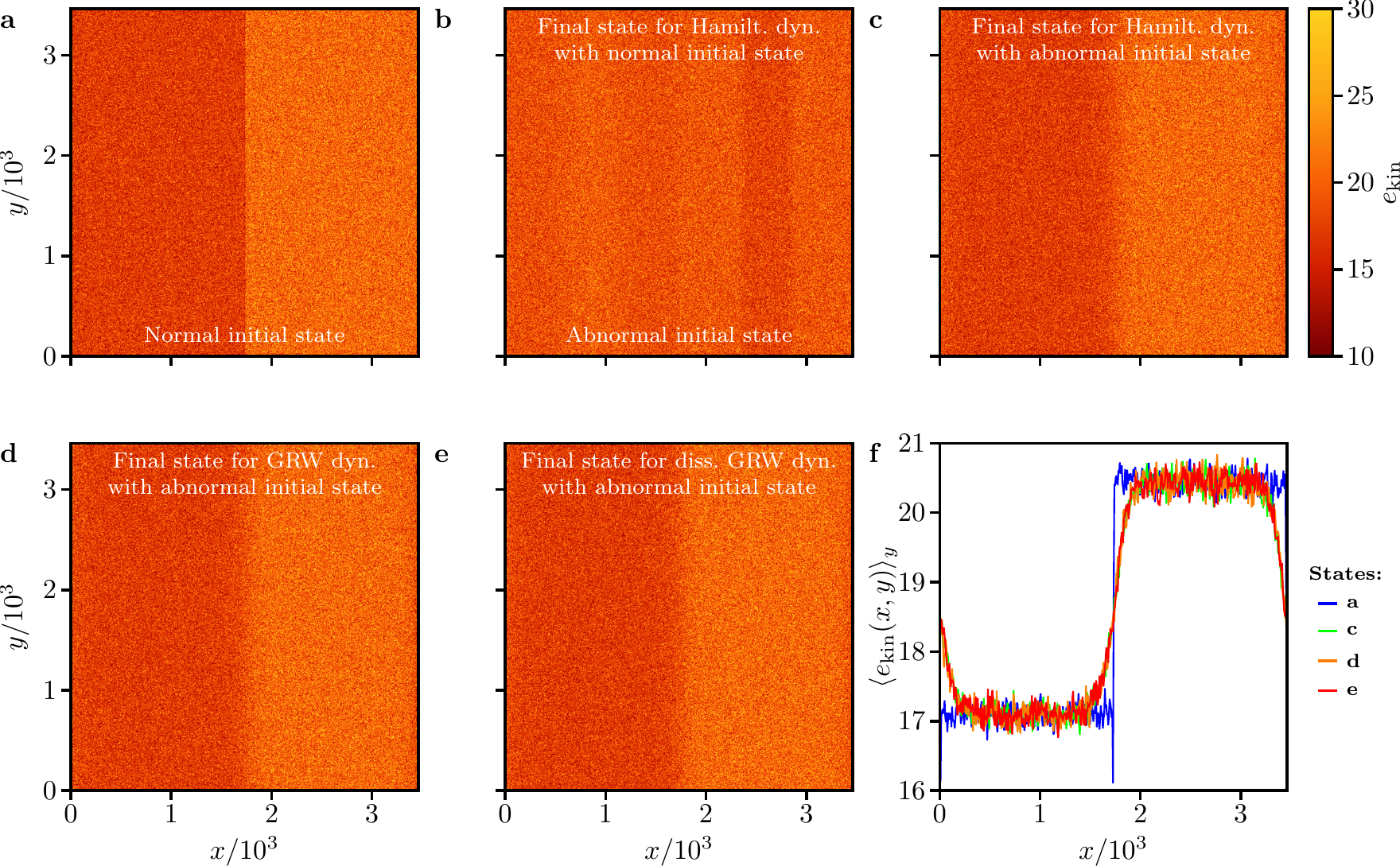}
\caption{\label{fig1}\textbf{(a)} Initial state of the first simulation (reference state). The system consists of two parts in thermal contact, separated by a sharp temperature gradient. \textbf{(b)} Final (relatively homogeneous) temperature distribution of the first simulation, obtained from evolving the system shown in (a) forward in time using Hamiltonian dynamics. \textbf{(c)}  Final temperature distribution of the second simulation, obtained from evolving a system with \ZT{abnormal} initial conditions (state shown in (b) with reversed momenta) under Hamiltonian dynamics. The system moves away from thermal equilibrium. \textbf{(d)} Final temperature distribution of the third simulation, obtained from evolving a system with the same initial condition as in (c) using GRW dynamics. The stochastic perturbations make no difference for the anti-thermodynamic behavior. \textbf{(e)} Final temperature distribution of the fourth simulation, obtained from evolving a system with the same initial condition as in (c) using dissipative GRW dynamics. This also leads to anti-thermodynamic behavior. \textbf{(f)} Comparison of the final temperature distributions of the second (deterministic), third (undamped stochastic), and fourth (damped stochastic) simulation with the initial state of the first one.}
\end{figure*}
To test Albert's proposal, we have performed simulations of a many-particle quantum system with and without GRW collapses. A first aspect to take into account here is the fact that our dynamics should allow for equilibration. This means that it is not possible to use an ideal gas, where \ZT{ideal} means that there are no interactions, not even collisions. Such an ideal gas is not guaranteed to approach equilibrium, since all particles will just move on a straight line without affecting each other. As soon as interactions are present, it is not possible to describe the system using a one-particle Wigner function without a closure for the interaction part of the dynamics. However, such closures often already introduce prior assumptions regarding the approach to equilibrium (such as the time-reversal symmetry breaking associated with the hypothesis of molecular chaos, see \cref{irre}), which we want to avoid here. Thus, we would have to solve the dynamic equation \eqref{generalgrwmodel} or \eqref{grwfokkermany} numerically for a many-particle system over a significant timescale, which is impossible. 

We can simplify the problem by noting that we are mainly interested in the effect of GRW collapses on the approach to equilibrium, and not in quantum effects such as superpositions. Using the Wigner framework, we have been able to derive the Fokker-Planck equation \eqref{grwfokkermany} which, although describing a quantum system, is very similar in form to classical Fokker-Planck equations. Thus, we can assume that we work in the classical limit where quantum effects apart from collapses (which, in the Fokker-Planck approximation, correspond to momentum diffusion) are negligible, such that particles have well-defined trajectories that can be described using Langevin equations.\footnote{Of course, this approximation ignores many important quantum effects, as discussed below. Nevertheless, our derivation shows that a classical Fokker-Planck equation, which is equivalent to a system of classical Langevin equations, can provide a good first approximation to incorporate GRW effects. In particular, it is important that Albert's approach is based simply on the fact that GRW collapses serve as a source for microscopic random perturbations rather than on details of the quantum-mechanical dynamics. Random perturbations are already well described by Langevin equations.} The Langevin equations corresponding to the Fokker-Planck equation \eqref{grwfokkermany} read
\begin{align}
\tdif{}{t}\vec{r}_i(t)&= \frac{\vec{p}_i}{m},\label{langevin1}\\
\tdif{}{t}\vec{p}_i(t)&= - \Nabla_{\vec{r}_i}U(\{\vec{r}_j\}) + \vec{\chi}_i(t),\label{langevin2}
\end{align}
where $\vec{\chi}_i(t)$ is white noise with the properties
\begin{align}
\braket{\vec{\chi}_i(t)}_\mathrm{E}&=\vec{0},\label{noise1}\\
\braket{\vec{\chi}_i(t)\otimes\vec{\chi}_j(t')}_\mathrm{E}&=2 D_p \Eins\delta_{ij}\delta(t-t')\label{noise2}
\end{align}
with the ensemble average $\braket{\cdot}_\mathrm{E}$, dyadic product $\otimes$, and identity matrix $\Eins$. Thus, we can base our simulations on the Langevin equations \eqref{langevin1} and \eqref{langevin2}. These are numerically much easier to handle than the many-particle Wigner function and thus allow us to model a large system. Moreover, this approach is in the spirit of Albert's discussion, which is based on the idea that the role of quantum collapses is to provide random perturbations, here represented by the noise term in \cref{langevin2}. If we use a dissipative model such as \cref{grwkramers} instead, \cref{langevin2} has to be replaced by
\begin{equation}
\tdif{}{t}\vec{p}_i(t)= -\gamma \vec{p}_i - \Nabla_{\vec{r}_i}U(\{\vec{r}_j\}) + \vec{\chi}_i(t)    
\label{langevin2damp}
\end{equation}
and the diffusion constant $D_p$ in \cref{noise2} is given by $D_p = \gamma m k_{\mathrm{B}} T_\mathrm{n}$. This then recovers the standard equations of motion for an underdamped Brownian particle, from which the GRW-based model given by \cref{langevin1,langevin2} can be obtained by taking the limit $\gamma \to 0$ at fixed $D_p$. We use a smoothed Lennard-Jones potential for the particle interactions (see \cref{smooth}). For our simulations, \cref{langevin1,langevin2} are nondimensionalized (see \cref{details}) and then solved using the Leapfrog algorithm with time step size $\dif t = 0.0025$. 

To see how our simulations should be set up, we take into account what precisely Albert's suggestion is based on: Although, if the initial state of the system is chosen randomly, it is much more likely that the system evolves towards equilibrium, there are certain (\ZT{abnormal}) initial conditions for which the system does \textit{not} evolve towards equilibrium. It is then the (expected) effect of the GRW collapses that the system is brought back onto the right track.

To see whether this actually works, we first require such an abnormal initial condition. For this purpose, we consider a many-particle system whose time evolution is governed by \cref{lvnwigner} (i.e., by unitary quantum mechanics without spontaneous collapses). Starting from an initial nonequilibrium distribution at $t=0$, we study its time evolution by solving the unitary dynamics (in its Newtonian approximation given by \cref{langevin1,langevin2} with $\vec{\chi} = \vec{0}$) numerically. After a certain time $t_\mathrm{rev}$, the system has evolved towards a distribution that is relatively close to an equilibrium state. We then stop and use the distribution at time $t=t_\mathrm{rev}$ with reversed momenta as the initial condition for a second simulation. By the time-reversal invariance of quantum and classical mechanics, the system should then evolve back towards the initial distribution. Since it has thereby moved from a close-to-equilibrium to a far-from-equilibrium state, we have thus found an \ZT{abnormal} initial condition (distribution at $t=t_\mathrm{rev}$ with reversed momenta) with the property that the system spontaneously moves away from equilibrium. 

Next, we perform another simulation with the same abnormal initial condition, but this time using GRW theory. As is obvious from \cref{grwfokkermany}, the dynamics with spontaneous collapse is no longer time-reversal invariant (the left-hand side and the first two terms on the right-hand side change signs under $(t,p)\to (-t,-p)$, the third term on the right-hand side does not). If spontaneous collapses can enforce the validity of the second law of thermodynamics in the way suggested by Albert, then the system should be observed to evolve towards equilibrium rather than away from it this time. 

The results are shown in \cref{fig1}, visualizing the local temperature measured by the kinetic energy density $e_{\mathrm{kin}}$ (see \cref{details}, all variables are dimensionless). We consider a dense fluid (the dimensionless\footnote{From the dimensionless number density $\varrho$, the dimensional number density can be obtained by multiplication with $\sigma^2$, where $\sigma$ is a length scale (see \cref{details}).} number density is $\varrho=0.7$), which makes the simulations more efficient and therefore allows us to see all relevant effects. For the first simulation, we prepare two two-dimensional systems of the same size and let them equilibrate at different temperatures $T_1 = 0.5335$ and $T_2=0.6391$.\footnote{We have set the temperatures in such a way that (a) the system behaves as a liquid, and (b) the two temperatures are different but not too far from each other (to avoid strongly nonlinear effects). The values of $T_1$ and $T_2$ are then obtained as the long-time averages in the equilibration simulations. These simulations had as an initial condition a homogeneous density with some noise and a Rayleigh velocity magnitude distribution. The standard deviation for the Gaussians used for the initial velocity distributions were 0.75 and 0.85, respectively.} Afterwards, the systems are brought in thermal contact. This reference state, which has a sharp temperature gradient, is shown in \cref{fig1}a. The total system contains about $8.5\cdot 10^6$ particles. We consider a quadratic domain with boundary length $L=3461.75$.\footnote{We have made the system as large as reasonably possible given the computational constraints. The precise particle number was then $N=2^{23}=8388608$, from which the domain length $L$ is obtained as $L=\sqrt{N/\varrho} \approx 3461.75$.} If we evolve the system forward in time following standard Hamiltonian dynamics (\cref{langevin1,langevin2} with $\vec{\chi}(t) = \vec{0}$) up to time $t_\mathrm{rev}=625$,\footnote{This time was chosen based on the available computing time. It is sufficiently long to achieve a reasonable degree of (but not full) equilibration, which is what we require for the problem at hand.} we observe -- in agreement with what one would expect from thermodynamics -- an equilibration towards a state with a more homogeneous temperature distribution. This relaxed state is shown in \cref{fig1}b. A more detailed analysis of this classical relaxation process can be found in Ref.\ \cite{Toth2020}. 

For the second simulation, we use as an initial condition the final state of the first one (shown in \cref{fig1}b), but with reversed particle momenta. This creates an \ZT{abnormal} initial condition that leads to anti-thermodynamic behavior under the Hamiltonian time evolution. The simulations confirm this: if we evolve this state forward in time using Hamilton's equations, we see that the system evolves from a state with a relatively homogeneous temperature (\cref{fig1}b) towards a state with a sharp temperature gradient shown in \cref{fig1}c, where the two subsystems have regained their initial temperatures. Consequently, the system has moved away from thermal equilibrium. This is surprising from a thermodynamic perspective, but should be expected from the fact that the underlying Hamiltonian microdynamics is time-reversible. Moreover, effects of this type are known from spin systems \cite{MicadeiEtAl2019,Hahn1950,RidderbosR1998}. 

Finally, Albert's suggestion is tested via the third simulation, which has the same initial condition as the second simulation, but uses the GRW-based model given by \cref{langevin1,langevin2} with nonvanishing noise $\vec{\chi}(t)$ (see \cref{details} for the noise amplitude). If his suggestion is correct, then the stochastic perturbations induced by the GRW dynamics should lead to \ZT{normal} thermodynamic behavior despite the abnormal initial condition. As can be seen from \cref{fig1}d, which shows the final state of the third simulation, this is not the case: the system moves away from thermal equilibrium towards a state with inhomogeneous temperature even in the presence of stochastic GRW-type perturbations. We have also performed a fourth simulation with the same initial condition as the second and third simulations, but using the dissipative GRW model given by \cref{langevin1,langevin2damp}. The final state of this simulation is shown in \cref{fig1}e, revealing that damping also does not lead to a restoration of thermodynamic behavior. Figure \ref{fig1}f, which compares the temperature profiles for the final states of the second (green), third (orange), and fourth (red) simulation to the initial state of the first simulation (blue), shows that the system evolves back to the initial state (with a small difference due to unavoidable numerical errors) regardless of whether or not stochastic GRW-type perturbations are present. Consequently, our results refute Albert's proposal according to which, for anti-thermodynamic initial conditions, thermodynamic behavior can be recovered via stochastic GRW-type perturbations.

The numerical experiments presented here are similar to the ones by \citet{OrbanB1967}, where anti-thermodynamic behavior was also achieved by time-reversing a molecular dynamics simulation. Orban and Bellemans found that the initial state is not completely recovered due to numerical round-off errors, from which they inferred that anti-thermodynamic behavior is unstable (similar in spirit to Albert's suggestion). Such \ZT{numerical irreversibility} has subsequently been investigated also by other authors (some of whom even linked it to quantum effects) \cite{KomatsuA2004,KomatsuA2005}. Since the effects of the round-off errors are now well understood, we can test what happens if GRW noise is added to the dynamics. Figure \ref{fig1}f shows that the difference between the reference state and the result of the deterministic time-reversed simulation (resulting purely from numerical errors) is larger than the difference between the deterministic and the stochastic simulations, even though the amplitude of the GRW noise is much larger than that of the \ZT{numerical noise} (such that our \ZT{null result} is not a consequence of numerical errors). The temperature gradient in the final state of the time-reversed simulations is smoother than in the reference state, suggesting that the way back from equilibrium has not been completed. However, this increased smoothness is present for both the deterministic and the stochastic simulations. This implies that the GRW noise has no significant effect on whether or not there is anti-thermodynamic behavior, and that the small differences between the initial state of the first simulation and the final states of the second and third simulations are a consequence of round-off errors.

\begin{figure*}[tbhp]
\centering\includegraphics[width=\linewidth]{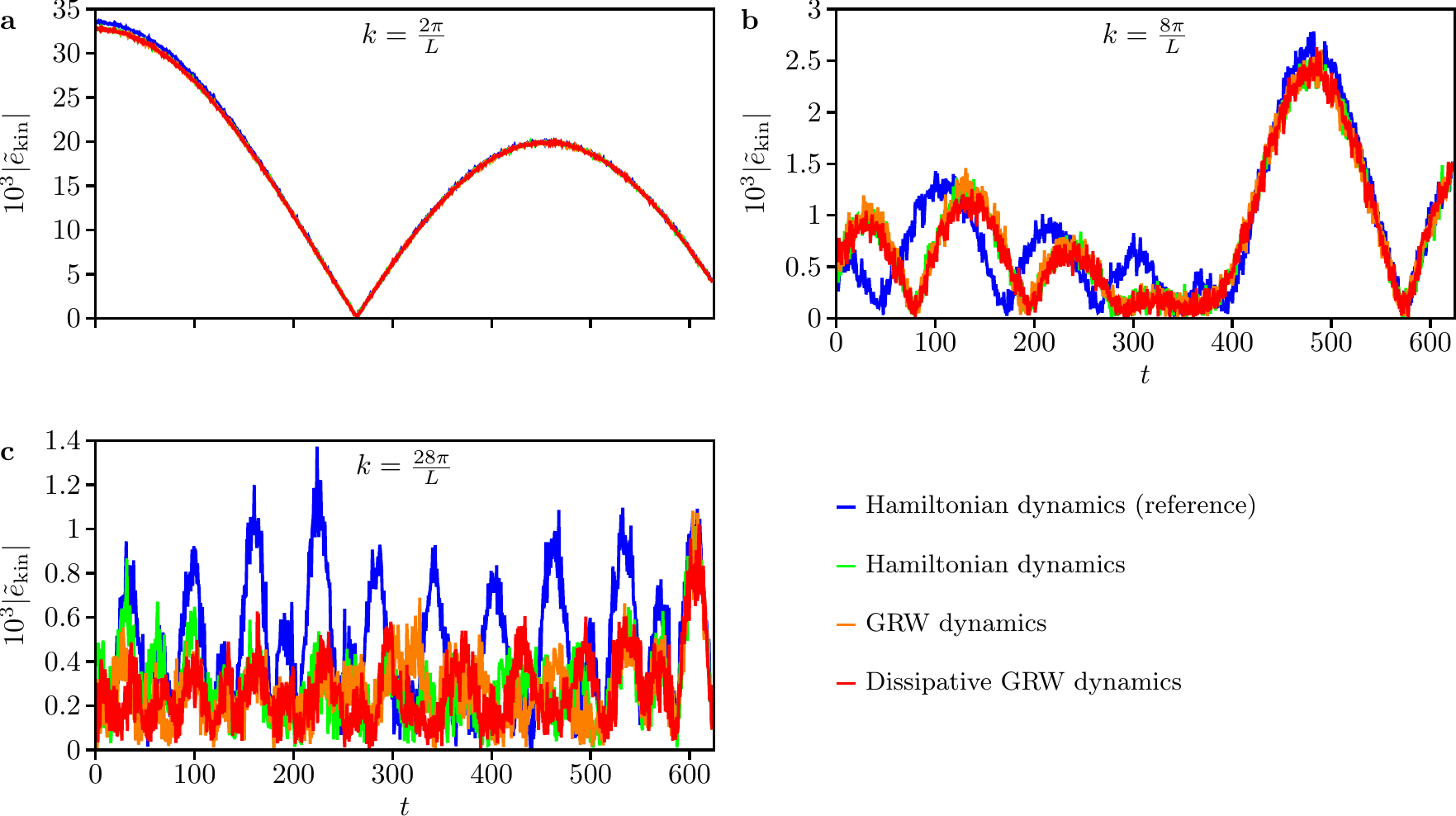}
\caption{\label{fig2}Time evolution of the absolute values of the Fourier modes of the kinetic energy density $|\tilde{e}_{\mathrm{kin}}|$ with wavenumbers \textbf{(a)} $k=2\pi/L$, \textbf{(b)} $k=8\pi/L$, and \textbf{(c)} $k=28\pi/L$ for the first (deterministic, reference), second (deterministic), third (undamped stochastic), and fourth (damped stochastic) simulation. For the second, third, and fourth simulation, time is reversed in this plot to allow for a comparison to the first simulation. The difference between deterministic and stochastic time-reversed dynamics (and between the different forms of stochastic dynamics) is larger for larger wavenumbers, but generally smaller than between the reference and the time-reversed simulations.}
\end{figure*}

A similar conclusion can be obtained from \cref{fig2}, which shows the time dependence of the absolute values of the Fourier modes of the kinetic energy density $|\tilde{e}_{\mathrm{kin}}|$ for the wavenumbers (a) $k=2\pi/L$, (b) $k=8\pi/L$, and (c) $k=28\pi/L$ for all three simulations (see \cref{details}, all variables are dimensionless). The second, third, and fourth simulations are plotted with reversed time to allow for a comparison to the first one. For all three wavenumbers, the agreement between the two time-reversed simulations is better than their agreement with the first (reference) simulation, implying that the GRW collapses make no significant difference for the macroscopic (thermodynamic) behavior of the system. Therefore, it is unlikely that they are responsible for the emergence of macroscopic irreversibility. However, the collapses do make a certain difference: as can be seen from comparing Figs.\ \ref{fig2}a and \ref{fig2}b to \cref{fig2}c, the disagreement between deterministic and stochastic simulation results (i.e., the effect of GRW noise) becomes larger for increasing wavenumber $k$. Since larger wavenumbers correspond to smaller length scales, this means that there are notable differences between deterministic and GRW dynamics on microscopic scales. However, these differences become less important on the macroscopic (thermodynamic) level. A similar result is found when comparing the damped and the undamped GRW model: there are differences on smaller length scales (larger wavenumbers), but not on the macroscopic level.

At this point, we should discuss three possible objections to our result. First, it can be argued that the relevant GRW parameters (in particular $D_p$) are so small that observable differences should be expected for macroscopic systems ($10^{23}$ particles), but not for systems with only a few particles (\citet{Albert1994} explicitly notes that, if his explanation is correct, an absolutely isolated gas with about $10^5$ particles would not be expected to show thermodynamic behavior). A macroscopic system would be far too large to be studied with microscopic simulations on reasonable timescales. To compensate for this fact, we have used a very strong noise. Assuming, for the sake of definiteness, that our fluid consists of argon atoms, our simulation parameters correspond to $D_p \approx 2.554 \cdot 10^{-43}$\,J$^2$s/m$^2$ (see \cref{details}), whereas using the parameter values $\lambda = 10^{-16}$/s and $\alpha = 10^{14}$\,m$^{-2}$ used in Ref.\ \cite{GhirardiRW1986} (to get an idea for the order of magnitude of $D_p$) gives (in two dimensions) $D_p = \lambda \alpha \hbar^2/2 \approx 5.56 \cdot 10^{-71}$\,J$^2$s/m$^2$. \citet{Adler2007} has suggested values of $\lambda$ that are about 10 orders of magnitude larger, increasing the possible value to $D_p \approx 10^{-61}$\,J$^2$s/m$^2$, which is still significantly below our value. If we assume that GRW effects become relevant if $\lambda N$ is of the order 1/s, then for $D_p \approx 10^{-61}$\,J$^2$s/m$^2$ (corresponding to $\lambda \approx 10^{-6}$/s) one would require about $10^6$ particles to see GRW effects. Since we have about $10^7$ particles and a noise that is about 9 orders of magnitude larger, there is no reason to assume that our \ZT{null result} for effects of GRW collapses on thermodynamics is a consequence of the system size. Current bounds for the GRW parameters are given in Ref.\ \cite{VinanteCBCVFMMU2020}. 

Second, we have tested a quantum-mechanical theory (GRW) using classical molecular dynamics simulations, motivated by the fact that solving the quantum-mechanical equations for a many-particle system is impossible. This can be justified using the Wigner formalism, which shows that if we ignore all quantum effects except for the spontaneous collapses, GRW theory is equivalent to a classical statistical theory for a system whose dynamics is governed by Langevin equations. Obviously, the result we would have obtained from a full quantum-mechanical calculation would be different. Nevertheless, if a full quantum-mechanical calculation would have led to a different result \textit{regarding the problem of irreversibility}, then irreversibility would not (solely) be a consequence of GRW collapses, but (also) of the other quantum effects (such as the peculiarities of quantum chaos) that have been ignored in our simulations. Clearly, for a system that is in a macroscopic superposition, the GRW collapses will make an important difference for the dynamics (namely, they will destroy this superposition). However, for explaining why a system of particles that already behave quasi-classically approaches thermodynamic equilibrium, they appear to play no significant role.

Third, we have made the approximation of moving from a master to a Fokker-Planck equation (e.g., from \cref{grwwigner} to \cref{grwfokker}) in the derivation of the Langevin equations \eqref{langevin1} and \eqref{langevin2}. This corresponds to the assumption that the Gaussian function that the Wigner function is convoluted with in \cref{grwwigner} decays very rapidly for larger values of $|p'|$. Since the full width at half maximum of this Gaussian is $2\sqrt{\ln(2)\alpha\hbar^2} \approx 1.756 \cdot 10^{-27}$\,kg\,m/s (which is extremely small), this assumption should also be unproblematic.

\section{\label{conclusion}Conclusion}
This article has two central results. First, we have derived master and Fokker-Planck equations for the Wigner function based on the GRW theory, the CSL model, the Di{\'o}si-Penrose model, and the dissipative GRW model. This provides a self-consistent dynamical theory in phase space that allows to explain the emergence of classicality from quantum mechanics regarding both the emergence of Liouville dynamics and the solution of the quantum measurement problem. Moreover, it makes the advantages of the Wigner framework available to researchers interested in spontaneous collapse models. Second, we have used Langevin equations derived from the GRW Fokker-Planck equation for a numerical test of Albert's suggestion that GRW collapses might be responsible for the approach to thermodynamic equilibrium. The simulations reveal that stochastic GRW-type perturbations do not lead to thermodynamic behavior if it is not already present in the deterministic dynamics. Consequently, our results do not support Albert's idea.

\acknowledgments{We thank Barbara Drossel, Julian Jeggle, Paul M. N\"ager, and David Wallace for helpful discussions. The simulations for this work were performed on the computer cluster PALMA II of the University of M\"unster.}

\section*{Funding}
M.t.V. thanks the Studienstiftung des deutschen Volkes for financial support. 
R.W.\ is funded by the Deutsche Forschungsgemeinschaft (DFG, German Research Foundation) -- WI 4170/3-1.

\section*{Declarations}
\subsection*{Data availability}
The datasets generated and/or analysed during the current study are available from the corresponding author on reasonable request.

\subsection*{Competing interests}
The authors have no relevant financial or non-financial interests to disclose.

\appendix
\section{\label{smooth}Interaction potential}
The potential in \cref{langevin2,langevin2damp} is (in nondimensionalized form) given by
\begin{equation}
U(\{\vec{r}_{k}\}) = \frac{1}{2}\underset{i \neq j}{\sum_{i,j=1}^{N}}U_2(\vec{r}_i - \vec{r}_j)
\end{equation}
with the pair-interaction potential $U_2$. For $U_2$, we use, following Ref.\ \cite{HammondsH2020}, the smoothed Lennard-Jones potential
\begin{equation}
U_2(r)  =
\begin{cases}
4 \big(\frac{1}{r^{12}} - \frac{1}{r^6}\big) + C_0 + C_2 r^2  & \\
+ C_4 r^4 + C_6 r^6 + C_8 r^8 &\text{ for } r \leq r_{\mathrm{c}},\\
0 &\text{ otherwise}
\end{cases}    
\label{smoothedlennard}
\end{equation}
with $r = \norm{\vec{r}}$, smoothing parameters $C_{0},\dotsc,C_{8}$, and cutoff length $r_{\mathrm{c}}$. We here use $r_c = 3.5$, implying \cite{HammondsH2020} $C_0 = 7.591 016 534 387 729 7 \cdot 10^{-2}$, $C_2 = -1.858 154 796 630 728 4 \cdot 10^{-2}$, $C_4 = 1.819 594 335 725 356 1 \cdot 10^{-3}$, $C_6 = -8.249 874 769 676 778 6 \cdot 10^{-5}$, and $C_8 = 1.442 811 390 098 910 1 \cdot 10^{-6}$. The form of the interaction potential and the choice of the smoothing parameters ensure that the potential and its first four derivatives vanish at $r = r_{\mathrm{c}}$.

\section{\label{details}Dimensionless quantities}
For the nondimensionalization of our equations, we first write \cref{langevin1,langevin2damp} as
\begin{equation}
m\tdif{^2}{t^2}\vec{r}_i(t) = -\gamma m \tdif{}{t}\vec{r}_i(t) - \Nabla_{\vec{r}_i}U(\{\vec{r}_j\}) + A \vec{\xi}_i(t), 
\end{equation}
where $A$ is the noise amplitude and $\vec{\xi}_i(t)$ has zero mean and correlation $\braket{\vec{\xi}_i(t)\otimes\vec{\xi}_j(t')}_\mathrm{E} = \Eins\delta_{ij}\delta(t-t')$. We then rescale $\vec{r}(t) = \sigma \vec{r}_\mathrm{nd}(t/\tau)$, $t = \tau t_\mathrm{nd}$, $U(\{\vec{r}_j\}) = \epsilon U_\mathrm{nd}(\{\vec{r}_j/\sigma\})$, $\vec{\xi}_i(t) = \vec{\xi}_{i,\mathrm{nd}}(t/\tau)/\sqrt{\tau}$, $\gamma = \gamma_\mathrm{nd}/\tau$, and $A =m\sigma A_\mathrm{nd}/\tau^{\frac{3}{2}}$, where $\sigma$ is a length scale, $\tau = \sigma\sqrt{\frac{m}{\epsilon}}$ a time scale, $\epsilon$ an energy scale, and the subscript nd denotes a nondimensionalized quantity. (This subscript is omitted in the main text and the figures.) We then obtain 
\begin{align}
\tdif{^2}{t^2}\vec{r}_{i,\mathrm{nd}}(t_{\mathrm{nd}}) &=- \gamma_\mathrm{nd}\tdif{}{t}\vec{r}_{i,\mathrm{nd}}(t_{\mathrm{nd}}) 
- \Nabla_{\vec{r}_{i,\mathrm{nd}}}U_\mathrm{nd}(\{\vec{r}_{j,\mathrm{nd}}\}) \notag\\ 
&\quad\,+ A_\mathrm{nd} \vec{\xi}_{i,\mathrm{nd}}(t_{\mathrm{nd}}).    
\label{nondimensionalized}
\end{align}

As discussed in \cref{numeric}, it is useful to compare the strength of the noise used in our simulations to the values expected from current estimates of the GRW model parameters. For the purposes of testing Albert's theory, it is good if the noise is \ZT{too large}, since we can thereby ensure that we are not overlooking any effects due to our \ZT{small} system size. An estimate of the order of magnitude is sufficient. The value of $D_p$ in SI units is given by
\begin{equation}
D_p= \frac{A_\mathrm{nd}^2\epsilon^2\tau}{2\sigma^2}.
\label{dp}
\end{equation}
In the Lennard-Jones potential, the parameters $\sigma$ and $\epsilon$ measure the range and depth of the potential, respectively. To get concrete numbers, we take the example of argon, for which the parameters $\sigma$ and $\epsilon$ take the values $\sigma = 3.4 \cdot 10^{-10}$\,m and $\epsilon \approx 1.657 \cdot 10^{-21}$\,J, respectively \cite{HansenMD2009}. Moreover, the mass of an argon atom is $m \approx 6.634 \cdot 10^{-26}$\,kg \cite{Lide2004}, such that $\tau \approx 2.151 \cdot 10^{-12}$\,s. From \cref{dp} and $A_\mathrm{nd} = 10^{-4}$ (value used in the simulations\footnote{To achieve a good description of GRW effects in approximately classical fluids, we have chosen the noise amplitude in such a way that it is significantly larger than the numerical round-off error, but also significantly smaller than the typical order of magnitude of the force resulting from the particle interactions.}), we then find $D_p \approx 2.554 \cdot 10^{-43}$\,J$^2$s/m$^2$. 

For the simulation with damping, we have used a noise amplitude $A_\mathrm{nd}=10^{-4}$ (as in the undamped case) and a noise temperature $T_\mathrm{n,nd}=0.5863$ related to the noise amplitude via the fluctuation-dissipation relation $A_\mathrm{nd}^2=2\gamma_\mathrm{nd}T_\mathrm{n,nd}$, which determines the dimensionless damping parameter $\gamma_\mathrm{nd}$. For the example of argon, this corresponds to a physical value $T_\mathrm{n} =T_\mathrm{n,nd} \epsilon/k_{\mathrm{B}}=70.356$\,K of the noise temperature.\footnote{To avoid over- or undercooling the system, the dimensionless noise temperature $T_\mathrm{n,nd}$ was simply chosen as the average of $T_1$ and $T_2$. This specific choice avoids the problem that (as discussed in \cref{irre}) the presence of the noise field will in general lead to a temperature different from the thermodynamically expected one (an effect that is also interesting, but different from the one we wish to investigate here). It is difficult to give an estimate of the real value of $T_\mathrm{n}$ (assuming that there is such a real value, i.e., that the dissipative GRW model is correct). Assuming that the noise has a cosmological origin, a value of about 1\,K is reasonable \cite{SmirneB2015}. Given that the damping parameter $\gamma$ is proportional to $\hbar^2/T_\mathrm{n}$ (see \cref{dgm}), this is already a \ZT{high} temperature in the sense that it leads to small damping. For our purposes, it is acceptable to choose a higher temperature since, due to the fluctuation-dissipation relation, this partially compensates for a too large value of $D_p$ when calculating $\gamma$.}

The dimensionless temperature $T_\mathrm{nd}$ (we here use $T_1 = 0.5335$ and $T_2=0.6391$) is defined as 
\begin{equation}
T_\mathrm{nd} = \frac{k_{\mathrm{B}} T}{\epsilon} = \frac{1}{dN}\sum_{i=1}^{N}\vec{v}_{i,\mathrm{nd}}^2 
\label{temperaturedefinition}
\end{equation}
with the dimensional temperature $T$, number of spatial dimensions $d$ (we use $d=2$ for the simulations), and dimensionless velocity $\vec{v}_{i,\mathrm{nd}}$ of the $i$-th particle. For argon, the temperatures used here correspond to $T_1\epsilon/k_{\mathrm{B}}= 64.02$\,K and $T_2 \epsilon/k_{\mathrm{B}}= 76.692$\,K.\footnote{This is below the freezing temperature for argon at normal conditions (83.8\,K) \cite{Lide2004}. Note, however, that the freezing temperature is affected by the density and that the potential \eqref{smoothedlennard} used here is not the exact interaction potential of argon. Nevertheless, since a Lennard-Jones fluid is typically a good model for argon \cite{HansenMD2009}, the estimates for the orders of magnitude of $D_p$ obtained for argon can be expected to be accurate.}

The \textit{local} temperature is measured by the local kinetic energy density
\begin{equation}
e_{\mathrm{kin}}(\vec{r}_{\mathrm{nd}}) = \frac{1}{2}\sum_{i=1}^{N}\vec{v}_{i,\mathrm{nd}}^2\delta(\vec{r}_{\mathrm{nd}}-\vec{r}_{i,\mathrm{nd}}).  
\end{equation}
Moreover, the $y$-averaged local kinetic energy density shown in \cref{fig1}d is given by $\braket{e_{\mathrm{kin}}}_y = \frac{1}{L}\TINT{}{}{y_\mathrm{nd}}e_{\mathrm{kin}}(\vec{r}_{\mathrm{nd}})$. For the figures, the local kinetic energy density was calculated via a mass redistribution technique\footnote{This technique, known as \ZT{cloud-in-cell method} \cite{BirdsallF1969}, is based on distributing the particle mass into four pieces located at adjacent nodes of a uniform quadratic grid. This is done in such a way that center of mass, momentum, angular momentum, and kinetic energy are conserved.}.

The Fourier coefficients shown in \cref{fig2} are defined as
\begin{equation}
\tilde{e}_{\mathrm{kin}}(\vec{k})=\frac{1}{2N}\sum_{i=1}^{N}\vec{v}_{i,\mathrm{nd}}^2e^{-\ii \vec{k}\cdot\vec{r}_{i,\mathrm{nd}}}, \label{fourierdefinition} 
\end{equation}
where $\vec{k}= (2\pi/L)\vec{n}$ with $\vec{n}\in \mathbb{N}^2$. The definition \eqref{fourierdefinition} ensures that $\tilde{e}_{\mathrm{kin}}(\vec{0})=T_\mathrm{nd}$ for $d=2$ (see \cref{temperaturedefinition}). In \cref{fig2}, we always set $k_y = 0$, such that $k = k_x = \norm{\vec{k}}$. 

\nocite{apsrev41Control}
\bibliographystyle{apsrev4-1}
\bibliography{refs,control}
\end{document}